\documentclass[11pt,a4paper,fleqn]{article}
\usepackage{amsmath}
\usepackage{amssymb}
\usepackage{amscd}
\usepackage{plain}
\usepackage{graphicx}
\usepackage{epsfig}
\usepackage{fancyhdr}
\usepackage{t1enc,epsfig}
\usepackage[mathscr]{eucal}
\usepackage{epstopdf}
\usepackage[german,english]{babel}

\newtheorem{theorem}{Theorem}[section]

\newtheorem{corollary}[theorem]{Corollary}
\newtheorem{definition}[theorem]{Definition}
\newtheorem{example}[theorem]{Example}
\newtheorem{lemma}[theorem]{Lemma}
\newtheorem{proposition}[theorem]{Proposition}
\newtheorem{remark}[theorem]{Remark}
\newtheorem{conjecture}[theorem]{Conjecture}
\numberwithin{equation}{section}

\newenvironment{proof}{{\bf Proof. }}{\hfill$\rule{1ex}{1ex}$\par\medskip}

\begin{document}

\newcommand{\bt}{\begin{theorem}}
\newcommand{\et}{\end{theorem}}
\newcommand{\bd}{\begin{definition}}
\newcommand{\ed}{\end{definition}}
\newcommand{\bs}{\begin{proposition}}
\newcommand{\es}{\end{proposition}}
\newcommand{\bp}{\begin{proof}}
\newcommand{\ep}{\end{proof}}
\newcommand{\be}{\begin{equation}}
\newcommand{\ee}{\end{equation}}
\newcommand{\ul}{\underline}
\newcommand{\br}{\begin{remark}}
\newcommand{\er}{\end{remark}}
\newcommand{\bex}{\begin{example}}
\newcommand{\eex}{\end{example}}
\newcommand{\bc}{\begin{corollary}}
\newcommand{\ec}{\end{corollary}}
\newcommand{\bl}{\begin{lemma}}
\newcommand{\el}{\end{lemma}}
\newcommand{\bj}{\begin{conjecture}}
\newcommand{\ej}{\end{conjecture}}

\def\phi{\varphi}
\def\epsilon{\varepsilon}

\def\ueta{\underline{\eta}}
\def\wt{\widetilde}
\def\fB{\mathfrak B}\def\fM{\mathfrak M}\def\fX{\mathfrak X}
 \def\cB{\mathcal B}\def\cM{\mathcal M}\def\cX{\mathcal X}
\def\mbe{\mathbf e}
\def\bu{\mathbf u}\def\bv{\mathbf v}\def\bx{\mathbf x} \def\by{\mathbf y} \def\bz{\mathbf z}
\def\om{\omega} \def\Om{\Omega}
\def\bbP{\mathbb P} \def\hw{h^{\rm w}} \def\hwi{{h^{\rm w}}}
\def\be{\begin{eqnarray}} \def\ee{\end{eqnarray}}
\def\beqq{\begin{eqnarray*}} \def\eeqq{\end{eqnarray*}}
\def\rd{{\rm d}} \def\Dwphi{{D^{\rm w}_\phi}}
\def\BX{\mathbf{X}}\def\Lam{\Lambda}\def\BY{\mathbf{Y}}
\def\BZ{\mathbf{Z}} \def\BN{\mathbf{N}}
\def\BV{\mathbf{V}}

\def\mwe{{D^{\rm w}_\phi}}
\def\DwPhi{{D^{\rm w}_\Phi}} \def\iw{i^{\rm w}_{\phi}}
\def\bE{\mathbb{E}}
\def\1{{\mathbf 1}} \def\fB{{\mathfrak B}}  \def\fM{{\mathfrak M}}
\def\diy{\displaystyle} \def\bbE{{\mathbb E}} \def\bu{\mathbf u}
\def\BC{{\mathbf C}} \def\lam{\lambda} \def\bbB{{\mathbb B}}
\def\bbR{{\mathbb R}}\def\bbS{{\mathbb S}}
 \def\bmu{{\mbox{\boldmath${\mu}$}}}
 \def\bPhi{{\mbox{\boldmath${\Phi}$}}}  \def\bPi{{\mbox{\boldmath{$\Pi$}}}}
  \def\btheta{{\mbox{\boldmath${\theta}$}}}
 \def\bbZ{{\mathbb Z}} \def\fF{\mathfrak F}\def\mbt{\mathbf t}\def\B1{\mathbf 1}
\def\hwphi{h^{\rm w}_{\phi}}
\def\BW{\mathbf{W}} \def\bw{\mathbf{w}}

\def\beal{\begin{array}{l}}
\def\beac{\begin{array}{c}}
\def\beacl{\begin{array}{cl}}
\def\ena{\end{array}}
\def\WBJ{\mathbf{J}^{\rm w}_{\phi}}
\def\BS{\mathbf{S}}
\def\BK{\mathbf{K}}
\def\BB{\mathbf{B}}
\def\wtD{{\widetilde D}}

\def\mwe{{D^{\rm w}_\phi}}
\def\DwPhi{{D^{\rm w}_\Phi}} \def\iw{i^{\rm w}_{\phi}}
\def\bE{\mathbb{E}}
\def\1{{\mathbf 1}} \def\fB{{\mathfrak B}}  \def\fM{{\mathfrak M}}
\def\diy{\displaystyle} \def\bbE{{\mathbb E}} \def\bu{\mathbf u}
\def\BC{{\mathbf C}} \def\lam{\lambda}
\def\bbB{{\mathbb B}} \def\bbM{{\mathbb M}}
\def\bbR{{\mathbb R}}\def\bbS{{\mathbb S}}
\def\blam{{\mbox{\boldmath${\lambda}$}}}
\def\bmu{{\mbox{\boldmath${\mu}$}}} \def\bta{{\mbox{\boldmath${\eta}$}}}
\def\bzeta{{\mbox{\boldmath${\zeta}$}}}
 \def\bPhi{{\mbox{\boldmath${\Phi}$}}}  \def\bPi{{\mbox{\boldmath{$\Pi$}}}}
 \def\bbZ{{\mathbb Z}} \def\fF{\mathfrak F}\def\mbt{\mathbf t}\def\B1{\mathbf 1}
\def\hwphi{h^{\rm w}_{\phi}}
\def\BT{{\mathbf T}} \def\BW{\mathbf{W}} \def\bw{\mathbf{w}}
\def\bfe{{\mathbf e}}
\def\beps{{\mathbf \varepsilon}}

\def\beal{\begin{array}{l}}
\def\beac{\begin{array}{c}}
\def\beacl{\begin{array}{cl}}
\def\ena{\end{array}}
\def\WBJ{\mathbf{J}^{\rm w}_{\phi}}
\def\BS{\mathbf{S}}
\def\BK{\mathbf{K}}
\def\tL{\mathbf{L}}
\def\BB{\mathbf{B}}
\def\vphi{{\varphi}}
\def\rw{{\rm w}}
\def\bZ{\mathbf Z}
\def\wtf{{\widetilde f}} \def\wtg{{\widetilde g}} \def\wtG{{\widetilde G}}
\def\vphi{\varphi}
\def\rT{{\rm T}}
\def\tA{{\tt A}} \def\tB{{\tt B}} \def\tC{{\tt C}} \def\tI{{\tt I}} \def\tJ{{\tt J}} \def\tK{{\tt K}}
\def\tL{{\tt L}} \def\tP{{\tt P}} \def\tQ{{\tt Q}} \def\tS{{\tt S}}
\def\beac{\begin{array}{c}} \def\beal{\begin{array}{l}} \def\beacl{\begin{array}{cl}} \def\ena{\end{array}}

\title{Basic inequalities for weighted entropies}

\author{Y. Suhov, I. Stuhl, S. Yasaei Sekeh, M. Kelbert}

\date{}
\footnotetext{\par}

\maketitle

\begin{abstract}
The concept of weighted entropy takes into account values of different outcomes, i.e., makes entropy  context-dependent, through the weight function. In this paper, we establish a number of simple inequalities for the weighted entropies (general as well as specific), mirroring similar bounds on standard (Shannon) entropies and related quantities. The required assumptions are written in terms of various expectations of the weight functions. Examples are weighted Ky Fan and weighted Hadamard inequalities involving determinants of positive-definite matrices, and weighted Cram\'{e}r-Rao inequalities involving the weighted Fisher information matrix.
\end{abstract}

\section{The weighted Gibbs inequality and its consequences}

The definition and initial results on weighted entropy were introduced in \cite{BG,G}. The purpose was to introduce disparity between outcomes of the same probability: in the case of a standard entropy such outcomes contribute the same amount of information/uncertainty, which is appropriate in context-free situations. However, imagine two equally rare medical conditions, occurring with probability $p<<1$, one of which carries a major health risk while the other is just a peculiarity. Formally, they provide the same amount of information $-\log\,p$ but the value of this information can be very different. The weight, or a weight function, was supposed to fulfill this task, at least to a certain extent.  The initial results have been further extended and deepened in \cite{ShMM,DT,KSa,PT,SiB,TCJ,K}, and, more recently, in \cite{DL,SrV,C,MMN,S}. Certain applications emerged, see \cite{FS1,JG}, along with a number of theoretical suggestions.

The purpose of this note is to extend a number of inequalities, established previously for a standard (Shannon) entropy, to the case of the weighted entropy. We particularly mention Ky Fan- and Hadamard-type inequalities from \cite{CT1, FS2, KS2} which are related to (standard) Gaussian entropies. Extended inequalities for weighted entropies already found applications and further developments in \cite{SuYSKel, SuYSSt, SuStKel}. Another kind of bounds, weighted Cram\'{e}r-Rao inequalities, may be useful in statistics.

An additional motivation for studying weighted entropy (WE) can be provided in the following questions. (I) What is the rate at which the WE is produced by a sample of a random process (and what could be an analog of the Shannon--McMillan--Breiman theorem)? (II) What would be an analog of Shannon's Second Coding theorem when an incorrect channel output causes a penalty but does not make the transmission session invalid? Properties of the WE established in the current paper could be helpful in this line of research.

One of naturally emerging questions is about the form/structure of the weight function (WF). In this paper we focus on some simple inequalities (as suggested by the title). Our results hold for fairly general WFs, subject to some mild conditions (in the form of inequalities). A systematic verification of these conditions may require a separate work.

Let $(\Om ,\fB,\bbP)$ be a standard probability space (see, e.g., \cite{I}). We consider random variables (RVs) as (measurable) functions $\Om\to\cX$, with values in
a measurable space $(\cX,\fM )$ equipped with a countably additive reference measure $\nu$.
Probability mass functions (PMFs) or probability density functions (PDFs) are denoted by
letter $f$ with various indices and defined
relative to $\nu$. The difference between PMFs (discrete parts of probability measures) and PDFs (continuous parts) is insignificant for most of the presentation;
this will be reflected in a common acronym PM/DF. In a few cases we will address
directly the probabilities $\bbP (X=i)$ (when $\cX$ is a finite or countable set, assuming that $\nu (i)=1$
$\forall$ $i\in\cX$). On the other hand,
some important facts will remain true without assumption that $\diy\int_{\cX}f(x)\nu (\rd x)=1$. When we
deal with a collection of RVs $X_i$, the space of values
$\cX_i$ and the reference measure $\nu_i$ may vary with $i$.
Some of RVs $X_i$ may be random $1\times n$ vectors, viz., $\BX_1^n=(X_1,\dots,X_n)$, with random
components $X_i:\Om\to\cX_i$, $1\leq i\leq n$.

\bd Given a function $x\in\cX\mapsto\phi (x )\geq 0$, and an RV
$X:\;\Om\to\cX$, with a PM/DF $f$,
the {\bf weighted entropy} (WE ) of $X$ (or $f$) with  weight function (WF) $\phi$ and
reference measure $\nu$ is defined by
\be\label{eq:1.01}
\hwphi (X)=\hwphi (f) =-\bbE[\phi (X)\log\,f(X)] = -\int_\cX\phi (x )f(x )\log\,f(x)\nu (\rd x )\ee
whenever the integral $\int_\cX\phi (x )f(x )\Big(1\vee|\log\,f(x)|\Big)\nu (\rd x )<\infty$. (A
standard agreement $0=0\cdot\log\,0=0\cdot\log\,\infty$ is adopted throughout the paper.)
If $f(x)\leq 1$ $\forall$ $x\in\cX$, $\hwphi (f)$ is non-negative. (This is the case when $\nu(\cX)\leq 1$.) The dependence of $\hwphi (X)=\hwphi (f)$
on $\nu$ is omitted.

Given two functions,  $x\in\cX\mapsto f(x )\geq 0$ and $x\in\cX\mapsto g(x )\geq 0$,
the {\bf relative} WE of $g$ relative to $f$ with WF $\phi$ is defined by
\be\label{eq:1.02}
\Dwphi (f\|g)=\int_\cX\phi (x )f(x )\log\frac{f(x )}{g(x )}\nu (\rd x ).
\ee
Alternatively, the quantity $\Dwphi (f\|g)$ can be termed a weighted
Kullback--Leibler divergence (of $g$ from $f$) with WF $\phi$. If $f$ is a PM/DF,
one can use an alternative form of writing: $$\Dwphi (f\|g)=\bbE\Big[\phi (X)\log\diy
\frac{f(X)}{g(X)}\Big].$$
\ed

In what follows, all WFs are assumed non-negative and positive on a set of positive $f$-measure.

\br
Passing to standard entropies, an obvious formula reads
\be
\hwphi (f)= h(\phi f) + D(\phi f \| f)=-D(\phi f\|\phi),
\ee
provided that one can guarantee that the integrals involved converge.
However, in general neither $\phi f$ nor $\phi$ are PM/DFs, which can be a nuisance. Besides,
the interpretation of $\phi$ as a weight function in $\hwphi (f)$ makes the inequalities more transparent.
\er

\bt\label{thm:1.1} {\rm{(The weighted Gibbs inequality; cf. \cite{CT2}, Lemma 1,  \cite{CT1}, Theorem 2.6.3, \cite{DCT} Lemma 1, \cite{KS2}, Theorem 1.2.3 (c).)}}
Given non-negative functions $f$, $g$, assume the bound
\be\label{eq:1.03}\int\limits_\cX\phi (x )\big[f(x)-g(x)\big]\nu (\rd x)\geq 0.\ee
Then
 \be\label{eq:1.03A}\Dwphi(f\|g)\geq 0.\ee
 Moreover, equality in \eqref{eq:1.03A} holds iff the ratio $\diy\frac{g}{f}$ equals $1$ modulo function $\phi$. In other words,
 $\left[\diy\frac{g(x)}{f(x)}-1\right]\phi (x)=0$ for $f$-almost all $x\in\cX$.
\et

\bp Following a standard calculation (see, e.g., \cite{CT1}, Theorem 2.6.3 or \cite{KS2}, Theorem 1.2.3 (c)) and using (\ref{eq:1.02}),
we write
\be\label{eq:1.04}\begin{array}{l}
-\Dwphi (f\|g)\;=\;\diy\int_\cX\phi (x)f(x)\1(f(x)>0)\log\frac{g(x)}{f(x)}\nu (\rd x)\\
\qquad\leq\diy \int_\cX\phi (x)f(x)\1(f(x)>0)\left[\frac{g(x)}{f(x)}-1\right]\nu (\rd x)\\
\qquad =\diy\int_\cX\phi (x) \1 (f(x)>0)\big[g(x)- f(x)\big]\nu (\rd x )
\leq \diy\int_\cX\phi (x)\big[g(x)-f(x)\big]\nu(\rd x)\leq 0.
\end{array}\ee
The equality in \eqref{eq:1.04} occurs iff $\phi (g/f- 1)$ vanishes $f$-a.s.
\ep

\bt\label{thm:1.2}
{\rm{(Bounding the WE via a uniform distribution.)}} Suppose an RV $X$ takes at most $m$ values, i.e.,
$\cX=\{1,\ldots ,m\}$, and set $p_i=\bbP (X=i)$, $1\leq i\leq m$. Suppose that for given $0<\beta\leq 1$
\be\label{ineq:1.1}\sum\limits_{i=1}^m\phi (i )(p_i-\beta ) \geq 0.\ee
Then $\hwphi(X)=-\sum\limits_{i=1}^m\phi (i)p_i\log\,p_i$ obeys
\be\label{eq:1.04A} \hwphi(X) \leq -\log\,\beta\sum\limits_{i=1}^m\phi (i)p_i,\;\;\;
\hbox{ or }\;\;\;-\bbE [\phi (X)\log p_X]\leq -(\log\,\beta )\bbE[\phi (X)], \ee
with equality iff for all $i=1,\dots,m$, $\phi (i )(p_i-\beta )=0$.

In the case of a general space $\cX$, assume that for a constant $\beta >0$ we have
\be\label{ineq1:2}\diy\int_\cX\phi (x)\left[f(x)-\beta\right]\nu (\rd x)\geq 0.\ee
Then
\be\label{eq:1.05} \hwphi (X)\leq-\log\,\beta\int_\cX\phi (x) f(x)\nu (\rd x);\ee
equality iff $\phi (x)\left[f(x)-\diy\beta\right]=0$ for
$f$-almost all $x\in\cX$.
\et

\bp The proof follows directly from Theorem \ref{thm:1.1}, with $g(x)=\beta$, $x\in\cX$.
\ep

\bd \label{Def multivarite}
Let $(X_1,X_2)$ be a pair of RVs $X_i:\;\Om\to\cX_i$, with a joint PM/DF
$f(x_1,x_2)$, $x_i\in\cX_i$,  $i=1,2$, relative to measure $\nu_1(\rd x_1)\times\nu_2(\rd x_2)$,
and marginal PM/DFs
$$f_1(x_1)=\diy\int_{\cX_2}f(x_1,x_2)\nu_2(\rd x_2), \;x_1\in\cX_1,\;\;f_2(x_2)=\diy\int_{\cX_1}f(x_1,x_2)\nu_1
(\rd x_1), \;x_2\in\cX_2.$$
Let $(x_1,x_2)\in\cX_1\times\cX_2\mapsto\phi (x_1,x_2)$ be a given WF. We use Eqn
\eqref{eq:1.01} to define the {\bf joint} WE of $X_1,X_2$ with WF $\phi$ (under an assumption of absolute
 convergence of the integrals involved):
\be\label{eq:1.06}\beacl\hwphi (X_1,X_2)&=-\bbE[\phi (X_1,X_2)\log f(X_1,X_2)]\\
\;&\diy=-\int_{\cX_1\times\cX_2}\phi (x_1,x_2 )f(x_1,x_2)\log\,
f(x_1,x_2)\nu_1(\rd x_1 )\nu_2(\rd x_2).\ena\ee
Next, the {\bf conditional} WE of $X_1$ given $X_2$ with WF $\phi$ is defined by
\be\label{eq:1.07}\beacl\hw_\phi (X_1|X_2)&\diy=-\bbE \Big[\phi (X_1,X_2)\log\,\frac{f(X_1,X_2)}{f_2(X_2)}
\Big]= \hw_\phi(X_1, X_2) - \hw_{\psi_{2}}(X_2)\\
\;&\diy=-\int_{\cX_1\times\cX_2}\phi (x_1,x_2)f(x_1,x_2)
\log\,\frac{f(x_1,x_2)}{f_2(x_2)}\nu_1(\rd x_1)\nu_2(\rd x_2),\ena\ee
here and below $$\psi_{2}(X_2)= \int_{\cX_1}\phi(x_1,x_2)\frac{f(x_1,x_2)}{f_2(x_2)}\nu_1(\rd x_1 ).$$
Further, the {\bf mutual} WE between $X_1$ and $X_2$ by
\be\label{eq:1.08} \beacl
\iw(X_1:X_2)=\mwe (f\|f_1\otimes f_2)=\diy \bbE \Big[\phi (X_1,X_2)\log\,\frac{f(X_1,X_2)}{f_1(X_1)f_2(X_2)}
\Big]\\
=\diy\int_{\cX_1\times\cX_2}\phi (x_1,x_2)f(x_1,x_2)\log\diy\frac{f(x_1,x_2)}
{f_1(x_1) f_2(x_2)}\nu_1(\rd x_1)\nu_2(\rd x_2).\ena\ee
\ed

We will use the notation $\BX_i^k=(X_i,\ldots ,X_k)$ and $\bx_i^k=(x_i,\ldots ,x_k)$, $1\leq i<k\leq n$,
for collections of RVs and their sample values (particularly for pairs and triples of RVs)
allowing us to shorten equations throughout the
paper. In addition, we employ Cartesian products  $\cX_i^k=\cX_i\times\ldots\times\cX_k$ and
product-measures $\nu_i^k(\rd\bx_i^k)=
\nu_i(\rd x_i)\times\ldots\times\nu_k(\rd x_k)$. Given a random $1\times n$ vector $\BX_1^n$ with a PM/DF $f$,
we denote by $f_i$,
$f_{ij}$ and $f_{ijk}$ the PM/DFs for component $X_i$, pair $\BX_{ij}=(X_i,X_j)$ and triple
$\BX_{ijk}=(X_i,X_j,X_k)$, respectively. The arguments of $f_i$, $f_{ij}$ and $f_{ijk}$ are written as
$x_i\in\cX_i$,
$\bx_{ij}=(x_i,x_j)\in\cX_{ij}=\cX_i\times\cX_j$ and $\bx_{ijk}=(x_i,x_j,x_k)\in\cX_{ijk}=\cX_i\times\cX_j
\times\cX_k$.
Next, symbols $f_{i|j}$, $f_{ij|k}$ and $f_{i|jk}$ are used for conditional PM/DFs:
$$f_{i|j}(x_i|x_j)=\frac{f_{ij}(\bx_{ij})}{f_j(x_j)},\;\;f_{ij|k}(\bx_{ij}|x_k)=\frac{f_{ijk}(\bx_{ijk})}{f_k(x_k)},\;\;
f_{i|jk}(x_i|\bx_{jk})=\frac{f_{ijk}(\bx_{ijk})}{f_{jk}(\bx_{jk})}.$$

For a pair of RVs $\BX_1^2$, set
\be\label{eq:1.09}\psi _1(x_1)=\diy\int_{\cX_2}\phi (x_1,x_2 )f_{2|1}(x_2|x_1)\nu_2(\rd x_2),\;x_1\in\cX_1;\ee
quantity $\psi _2(x_2)$, $x_2\in\cX_2$, is defined in a similar (symmetric) fashion. See above.

Next, given a triple of RVs $\BX_1^3$, with a joint PM/DF $f(\bx_1^3)$, set:
\be\label{eq:1.15}\begin{array}{c}\psi_3^{12}(x_3)=\diy\int_{\cX_1^2}\phi (\bx_1^3)f_{12|3}
(\bx_1^2|x_3)\nu_1^2(\rd\bx_1^2)=\bbE\Big[\phi (\BX_1^3)|X_3=x_3\Big],\;x_3\in\cX_3,\\
\psi_{12}(\bx_1^2)=\diy\int_{\cX_3}\phi (\bx_1^3)f_{3|12}(x_3|\bx_1^2)\nu_3(\rd x_3)
=\bbE\Big[\phi (\BX_1^3)|\BX_1^2=\bx_1^2\Big],\;\bx_1^2\in\cX_1^2,\end{array}\ee
and define functions $\psi_k^{ij}$ and $\psi_{ij}$ for distinct labels $1\leq i,j,k\leq 3$, in a similar manner.

\bl\label{lem:1.1} {\rm{(Bounds on conditional WE, I.)}} Let $\BX_1^2$
be a pair of RVs with a joint PM/DF $f(\bx_1^2)$. Suppose that a WF\; $\bx_1^2
\in\cX_1^2\mapsto\phi (\bx_1^2)$ obeys
\be\label{eq:1.10}\bbE\Big[\phi (\BX_1^2)\big[f_{1|2}(X_1|X_2)-1\big]\Big] =\int_{\cX_1^2}\phi (\bx_1^2)f(\bx_1^2)\Big[f_{1|2}(x_1|x_2)-1\Big]
\nu_1^2(\rd\bx_1^2)\leq 0.\ee
Then
\be\label{eq:1.11}\hwphi (\BX_1^2)\geq \hw_{\psi_2^{\,}}(X_2),\;\hbox{ or, equivalently, }\; \hwphi (X_1|X_2)\geq 0,\ee
with equality iff \;$\phi (\bx_1^2)\big[f_{1|2}(x_1|x_2)-1\big]=0$\; for $f$-almost all $\bx_1^2\in\cX_1^2$.
\el

\bp The statement is derived similarly to Theorem \ref{thm:1.1}:
$$\int_{\cX_1^2}\phi (\bx_1^2)f(\bx_1^2)\log\,f_{1|2}(x_1|x_2)\nu_1^2(\rd\bx_1^2)
\leq\int_{\cX_1^2}\phi (\bx_1^2)f(\bx_1^2)\Big[f_{1|2}(x_1|x_2)-1\Big]\nu_1^2(\rd\bx_1^2).
$$
The argument is concluded as in \eqref{eq:1.04}. The cases of equalities also follow.
\ep

\br\label{rem1.7}
In particular, suppose that $X_1$ takes finitely or countably many values and $\nu_1$ is a counting measure
with $\nu_1(i)=1$, $i\in\cX_1$. Then the value $f_{1|2}(x_1|x_2)$ yields the conditional probability
$\bbP (X_1=x_1|x_2)$, which is $\;\leq 1$\; for \;$f_2$-almost all $x_2\in\cX_2$. Then $\hwphi (X_1|X_2)\geq 0$, and the bound is strict unless,
modulo $\phi$,
RV $X_1$ is a function of $X_2$. That is, there exists a map $\upsilon :\cX_2\to\cX_1$ such that
$\big[x_1-\upsilon (x_2)\big]\phi (\bx_1^2)=0$ for $f$-almost every $\bx_1^2\in\cX_1^2$.
\er

For a future use, we can consider a triple of RVs, $\BX_1^3$, and a pair,
$\BX_2^3$, and assume that
\be\label{eq:1.10A}\bbE \Big[\phi (\BX_1^3)\big[f_{1|23}(X_1|\BX_2^3)-1\big]\Big] =\int_{\cX_1^3}\phi (\bx_1^3)f(\bx_1^3)\Big[f_{1|23}(x_1|\bx_2^3)-1\Big]
\nu_1^3(\rd\bx_1^3)\leq 0.\ee
Then
\be\label{eq:1.11A}\hwphi (\BX_1^3)\geq \hw_{\psi_{23}^{\,}}(\BX_2^3),\;\hbox{ or, equivalently, }\; \hwphi (X_1|\BX_2^3)\geq 0,\ee
with equality iff \;$\phi (\bx_1^3)\big[f_{1|23}(x_1|\bx_2^3)-1\big]=0$\; for $f$-almost all $\bx_1^3\in\cX_1^3$.

\bt\label{thm:1.3}
{\rm{(Sub-additivity of the WE.)}} Let $\BX_1^2=(X_1,X_2)$ be a pair of
RVs  with a joint
PM/DF $f(\bx_1^2)$ and marginals $f_1(x_1)$, $f_2(x_2)$, where $\bx_1^2\in\cX_1^2$.
Suppose that a WF $\bx_1^2\in\cX_1^2\mapsto\phi (\bx_1^2)$ obeys
\be\label{eq:1.12}\bbE\phi (\BX_1^2) -\bbE \phi (\BX^{\otimes}_{12})
=\int_{\cX_1^2}\phi (\bx_1^2)\Big[f(\bx_1^2)-f_1(x_1)f_2(x_2)\Big]\nu_1^2
(\rd\bx_1^2)\geq 0.\ee
Here $\BX^{\otimes}_{12}$ stands for the pair of independent RVs having the same
marginal distributions as $X_1$, $X_2$. (The joint PDF for $\BX^{\otimes}_{12}$
is the product $f_1(x_1)f_2(x_2)$.)  Then
\be \label{eq:1.13}\begin{array}{c} \hwphi(\BX_1^2)\leq
\hw_{\psi _1}(X_1)+\hw_{\psi _2}(X_2),\;\hbox{ or, equivalently, }\;\hwphi (X_1|X_2)\leq
\hw_{\psi_1^{\,}}(X_1),\\
\;\hbox{ or, equivalently, }\;i^{\rm w}_\phi (X_1:X_2)\geq 0.\end{array}\ee
The equalities hold iff  $X_1$, $X_2$ are independent modulo $\phi$, i.e.,
$$\phi (\bx_1^2)\left[1-\diy\frac{f_1(x_1)
f_2(x_2)}{f(\bx_1^2)}\right]=0$$ for $f$-almost all $\bx_1^2\in\cX_1^2$.
\et

\bp The subsequent argument works for the proof of Theorem \ref{thm:1.4} as well.
Set \\
$\left(f_1\otimes f_2\right)(x_1,x_2)=f_1(x_1)f_2(x_2)$. According to \eqref{eq:1.02}, \eqref{eq:1.06}
-- \eqref{eq:1.08} and owing to Theorem \ref{thm:1.1} and Lemma \ref{lem:1.1},
\be\label{eq:1.14}\begin{array}{lll} 0\geq -\mwe (f\|f_1\otimes f_2)&=&\diy
\int_{\cX_1^2}\phi(\bx_1^2 )f(\bx_1^2)\log
\frac{f_1(x_1)f_2(x_2)}{f(\bx_1^2)}\nu_1^2(\rd\bx_1^2)\\
&=&\hw_\phi (X_1,X_2)-\hw_{\psi _1}(X_1)-\hw_{\psi _2}(X_2)\\
&=&\hw_\phi(X_1|X_2)-\hw_{\psi _1}(X_1)=-i^{\rm w}_\phi (X_1:X_2). \end{array}\ee
This yields the inequalities in \eqref{eq:1.13}. The cases of equality
are also identified from Theorem
\ref{thm:1.1}.
\ep

Note that if in \eqref{eq:1.12} we use function $\psi_{12}(\bx_1^2)$ emerging from triple $\BX_1^3$, the assumption becomes
\be\label{eq:1.12A}\beal\bbE\phi (\BX_1^3)-\bbE \phi (\BX_{12}^{\otimes}\to X_3)\\
\qquad \diy =\int_{\cX_1^3}\phi (\bx_1^3 )\Big[f_{12}(\bx_1^2)-f_1(x_1)f_2(x_2)\Big]
f_{3|12}(x_3|\bx_1^2)\nu_1^3 (\rd\bx_1^3)\geq 0\ena\ee
and the conclusion
\be \label{eq:1.13A}h^{\rm w}_{\psi_{12}^{\,}} (X_1|X_2)\leq
\hw_{\psi_1^{23}}(X_1).\ee
Here $\BX_{12}^{\otimes}\to X_3$ denotes the triple of RVs where $X_1$ and $X_2$
have been made independent, keeping intact their marginal distributions, and
$X_3$ has the same conditional PM/DF $f_{3|12}$ as within the original triple
$\BX_1^3$.

\bl\label{lem:1.2} {\rm{(Bounds on conditional WE, II.)}} Let  $\BX_1^3$ be
a triple of RVs, with a joint PM/DF $f(\bx_1^3)$.
Given  a WF $\bx_1^3\mapsto\phi (\bx_1^3)$,
assume that
\be\label{eq:1.16}\bbE\Big[\phi (\BX_1^3)\big[f_{1|23}(X_1|\BX_2^3)-1\big]\Big] =\int_{\cX_1^3}\phi (\bx_1^3)f(\bx_1^3)\Big[f_{1|23}(x_1|\bx_2^3)-1\Big]\nu_1^3(\rd\bx_1^3)
\leq 0.\ee
Then
\be \label{eq:1.17} \hw_{\psi _{23}^{\,}}(X_2|X_3)
\leq \hwphi(\BX_1^2|X_3);\ee
equality iff \;$\phi (\bx_1^3)\big[f_{1|23}(x_1|\bx_2^3)-1\big]=0$\; for $f$-almost all $\bx_1^3\in\cX_1^3$.

As in Remark {\rm\ref{rem1.7}}, assume $X_1$ takes finitely or countably many values and $\nu_1(i)=1$,
$i\in\cX_1$. Then the value $f_{1|23}(x_1|\bx_2^3)$ yields the conditional probability
$\bbP (X_1=x_1|\bx_2^3)$, for \;$f_{23}$-almost all $\bx_2^3\in\cX_2^3$. Then $\hwphi (\BX_1^2|X_3)\geq
\hw_{\psi _{23}^{\,}}(X_2|X_3)$, with equality iff
modulo $\phi$,
RV $X_1$ is a function of $\BX_2^3$.
\el

\bp Observe that $\hwphi (\BX_1^2|X_3)=\hwphi (\BX_1^3)-\hw_{\psi_3^{12}}(X_3)$ and
$\hw_{\psi_{23}^{\,}} (X_2|X_3)=\hw_{\psi_{23}^{\,}} (\BX_2^3)-\hw_{\psi_3^{12}}(X_3)$, so that we need
to prove that $\hwphi (\BX_1^3)\geq\hw_{\psi_{23}^{\,}} (\BX_2^3)$.  The proof follows that of Lemma
 \ref{lem:1.1}, with obvious modifications.
\ep

Of course, if we swap labels $1$ and $3$ in \eqref{eq:1.16}, assuming that
 \be\label{eq:1.18}
\bbE\phi (\BX_1^3)\Big[f_{3|12}(X_3|\BX_1^2)-1\Big] = \int_{\cX_1^3}\phi (\bx_1^3)f(\bx_1^3)\Big[f_{3|12}(x_3|\bx_1^2)-1\Big]\nu_1^3(\rd\bx_1^3)
\leq 0\ee
 we get
 $$\hw_{\psi _{12}^{\,}}(X_1|X_2)
\leq \hwphi(\BX_{13}|X_2),$$
with equality iff \;$\phi (\bx_1^3)\big[f_{3|12}(x_3|\bx_1^2)-1\big]=0$\; for $f$-almost all $\bx_1^3\in\cX_1^3$.

\bt\label{thm:1.4} {\rm{(Sub-additivity of the conditional WE.)}} Let  $\BX_1^3$ be
a triple of RVs, with a joint PM/DF $f$.
Given  a WF $\bx_1^3\mapsto\phi (\bx_1^3)$, assume the following bound
\be\label{eq:1.19}\beal\bbE\phi (\BX_1^3) -\bbE\phi (X_2\to \BX_{13}^\otimes )\\
\qquad\diy=\diy\int_{\cX_1^3}\phi (\bx_1^3)\left[f (\bx_1^3)-f_2(x_2)\prod\limits_{i=1,3}f_{i|2}(x_i|x_2)
\right]\nu_1^3(\rd\bx_1^3)\geq 0.\ena\ee
Here $X_2\to \BX_{13}^\otimes$ stands for the triple of RVs where
$X_2$ keeps its distribution as within the triple $\BX_1^3$ whereas $X_1$ and $X_3$ have been made conditionally independent given $X_2$, with the same marginal
conditional PDFs $f_{1|2}$ and $f_{3|2}$ as in $\BX_1^3$. Then
\be \label{eq:1.24} \hwphi(\BX_{13}|X_2)\leq
\hw_{\psi _{12}^{\,}}(X_1|X_2)+\hw_{\psi_{32}^{\,}}(X_3|X_2),\ee
with equality iff, modulo $\phi$, RVs  $X_1$ and $X_3$ are conditionally independent given $X_2$.
That is: $\phi (\bx_1^3)\Big[f(\bx_1^3)-f_2(x_2)f_{1|2}(x_1|x_2)f_{3|2}(x_3|x_2)\Big]=0$ for $f$-almost all
$\bx_1^3\in\cX_1^3$.
\et

\bp The proof is based on the equation \eqref{eq:1.25}:
\be\label{eq:1.25}\begin{array}{l} \hwphi (\BX_{13}|X_2)-\hw_{\psi_{12}^{\,}}(X_1|X_2)-\hw_{\psi_{32}^{\,}}(X_3|X_2)\\
\qquad =\diy\int_{\cX_1^3}\phi (\bx_1^3)f(\bx_1^3)\log\,\frac{f_{1|2}(x_1|x_2)f_{3|2}(x_3|x_2)}{f_{13|2}
(\bx_{13}|x_2)}\\
\qquad =\diy\int_{\cX_1^3}\phi (\bx_1^3)f(\bx_1^3)\log\,\frac{f_2(x_2)f_{1|2}(x_1|x_2)f_{3|2}(x_3|x_2)}{
f(\bx_1^3)}\,.\end{array}\ee
After that we apply the same argument as in \eqref{eq:1.14}.
\ep

\bl\label{lem:1.3} {\rm{(Bounds on conditional WE, III.)}}
For a triple of RVs $\BX_1^3$ with a joint PM/DF $f(\bx_1^3)$ and a WF\; $\bx_1^3
\mapsto\phi(\bx_1^3)$,  assume the bound as in {\rm{\eqref{eq:1.19}}}.
Then
\be \label{eq:1.26}\begin{array}{r} \hwphi(X_1|\BX_2^3)\leq \hw_{\psi _{12}^{\,}}(X_1|X_2);\;\hbox{
equality iff  $X_1$ and $X_3$}\qquad{}\\ \hbox{are conditionally independent given $X_2$ modulo $\phi$.}\end{array}\ee
 \el

\bp Write \eqref{eq:1.26}
as
$$h_{\psi_{12}^{\,}}(\BX_1^2)-h_{\psi_2^{13}}(X_2)\geq \hwphi (\BX_1^3)-\hw_{\psi_{23}}(\BX_2^3)$$
and then pass to an equivalent form $\hwphi (\BX_{13}|X_2)\leq
\hw_{\psi _{12}^{\,}}(X_1|X_2)+\hw_{\psi_{32}^{\,}}(X_3|X_2)$ which is exactly
(\ref{eq:1.24}).  \def\lam{{\lambda}}
\ep

Summarizing, we have an array of inequalities \eqref{eq:1.27} for $\hwphi (X_1|\BX_2^3)$ and its upper bounds, each
requiring its own assumption (and with its own case for equality):
\be\label{eq:1.27}\begin{array}{l}\;\;\;\,\hbox{by Lemma \ref{lem:1.1}:}\;\;\;\;\;\;\;0\leq\;\hwphi (X_1|\BX_2^3),
\;\hbox{assuming \eqref{eq:1.10A} (a modified form of \eqref{eq:1.10}),}\\
\;\;\diy\begin{array}{ll}
\hbox{by Lemma \ref{lem:1.3}:}&\hwphi (X_1|\BX_2^3)\leq \hw_{\psi_{12}}(X_1|X_2), \;
\hbox{assuming (\ref{eq:1.19}),}\\
\hbox{by Theorem \ref{thm:1.3}:}&\hw_{\psi_{12}}(X_1|X_2)\leq h_{\psi_1^{23}}(X_1),
 \;\hbox{assuming (\ref{eq:1.12A}),}\\
 \hbox{by Lemma \ref{lem:1.2}:}&\hw_{\psi_{12}^{\,}} (X_1|X_2)\leq \hw_\phi(\BX_{13}|X_2), \;
 \hbox{assuming \eqref{eq:1.18},}\\
\hbox{by Theorem \ref{thm:1.4}:} &\hw_\phi(\BX_{13}|X_2)\leq \hw_{\psi_{12}^{\,}}(X_1|X_2)
+\hw_{\psi_{32}^{\,}}(X_3|X_2),\;\hbox{assuming \eqref{eq:1.19}.}\end{array}\end{array}\ee
It is worth noting that the assumptions listed in Eqn \eqref{eq:1.27} express
an impact on the total expected weight when we perform various manipulations
with RVs forming a pair or a triple under consideration.

\bt\label{thm:1.5}
{\rm (Strong sub-additivity of the WE.)} Given a triple of RVs $\BX_1^3$, assume that
bound {\rm{(\ref{eq:1.19})}} is fulfilled. Then
\be\label{eq:1.28} \hwphi(\BX_1^3)+\hw_{\psi_2^{13}}(X_2)\leq \hw_{\psi_{12}^{\,}}(\BX_1^2)+
\hw_{\psi_{23}^{\,}}(\BX_2^3).\ee
The equality in \eqref{eq:1.28} holds iff, modulo $\phi$,  $X_1$ and $X_3$ are conditionally independent
given $X_2$.
\et

\def\hwPhi{{\hw_\Phi}}

\bp Write the inequality in Eqn \eqref{eq:1.28} in an equivalent form:
\be \label{eq:1.29}\hwphi(\BX_1^3)-\hw_{\psi_2^{13}}(X_2)\leq \hw_{\psi_{12}^{\,}}(\BX_1^2)
-\hw_{\psi_2^{13}}(X_2)+
\hw_{\psi_{23}^{\,}}(\BX_2^3)-\hw_{\psi_2^{13}}(X_2).\ee
The LHS in \eqref{eq:1.29} equals $\hw_\phi (\BX_{13}|X_2)$ while the RHS yields
$\hw_{\psi_{12}^{\,}}(X_1|X_2)+\hw_{\psi_{32}^{\,}}(X_3|X_2)$. The inequality then follows from
Theorem \ref{thm:1.4}.
\ep

\def\Tha{{\Theta}} \def\tha{{\theta}} \def\wt{\widetilde}

\section{Convexity, concavity, data-processing and Fano inequalities}

\bt\label{thm2-1}
{\rm (Concavity of the WE; cf. \cite{CT1}, Theorem 2.7.3.)} The functional $f\mapsto \hwphi(f)$ is concave in argument $f$. Namely,
for given PM/DFs $f_1(x)$, $f_2(x)$, non-negative function $x\in\cX\mapsto\phi (x)$
and $\lam_1,\lam_2\in [0,1]$ with $\lam_1+\lam_2=1$,
\be\label{eq:2.1}
\hwphi (\lam_1f_1+\lam_2f_2)\geq \lam_1\hwphi (f_1)+\lam_2\hwphi (f_2).\ee
The inequality in \eqref{eq:2.1} is strict unless one of the values $\lam_1$, $\lam_2$ vanishes (and the other
 equals $1$) or when $f_1$ and $f_2$ coincide modulo $\phi$, that is, $\phi (x)\big[f_1(x)-f_2(x)\big]=0$ for $(\lam_1f_1+\lam_2f_2)$-almost all $x\in\cX$.
\et

\bp
Let $X_1,X_2:\;\Om\to\cX$ be RVs with PM/DF $f_1$ and $f_2$, respectively.
Consider a binary RV $\Theta$ with
\be
\Theta= \begin{cases}
          1, &  \hbox{with  probability $\lam_1$,} \\
          2, &  \hbox{with probability $\lam_2$.} \end{cases}\ee
Setting $Z=X_\theta$ yields an RV $Z$ with values from $\cX$ and with PM/DF $f=\lambda_1 f_1+\lambda_2 f_2$. Thus,
$$\hwphi(Z)=\hwphi(\lambda_1 f_1+\lambda_2 f_2).$$
On the other hand, take the conditional WE $\hw_{\wt\phi}(Z|\Theta)$ with the WF $\wt\phi (z,\tha )=\phi (z)$ depending on the first argument $z\in\cX$ and not on value $\tha =1,2$ of RV $\Tha$. Then the WF $\psi_1(z)=\bbE\Big[ {\wt\phi}(Z,\Tha ) |Z=z\Big]$ coincides with $\phi (z)$. It means that condition \eqref{eq:1.12} hold true for the pair of RVs $Z, \Theta$. According to Theorem \ref{thm:1.3} (cf. Eqn \eqref{eq:1.13}),
$\hw_{\wt\phi}(Z|\Theta)\leq\hwphi (Z)$, with equality iff $Z$ and $\Tha$ are independent modulo $\phi$. The latter holds when the product $\lam_1\lam_2 =0$ or when $f_1=f_2$ modulo $\phi$. Now,
$$\hwphi(Z|\Theta)=-\sum\limits_{\theta=1}^2\lam_\theta \int_{\cX}\phi(z) f_\theta(z)\log \, f_\theta(z)\nu (\rd z)
=\lam_1\hwphi (f_1)+\lam_2\hwphi (f_2).$$
This completes the proof.
\ep

\bt \label{thm:2.2}{\rm{(a)}}
{\rm (Convexity of relative WE; cf. \cite{CT1}, Theorem 2.7.2.)} Consider two pairs of non-negative functions, $(f_1,g_1)$ and $(f_2,g_2)$, on $\cX$. Given
a WF $x\in\cX\mapsto\phi (x)$ and $\lambda_1\lambda_2 \in(0,1)$ with $\lambda_1+\lambda_2=1$, the
following property is satisfied:
\be\label{eq:2.2} \lambda_1 \Dwphi(f_1\|g_1)+\lambda_2\Dwphi(f_2\|g_2)\geq \Dwphi(\lambda_1f_1+
\lambda_2 f_2\|\lambda_1g_1+\lambda_2g_2),\ee
with equality iff $\lam_1\lam_2 =0$ or $f_1=f_2$ and $g_1=g_2$ modulo $\phi$.

{\rm{(b)}} {\rm{(Data-processing inequality for relative WE; cf. \cite{CT1}, Theorem 2.8.1.)}} Let $(f,g)$ be a pair of non-negative functions
and $\phi$
a WF on $\cX$. Let  ${\mbox{\boldmath${\Pi}$}}=(\Pi (x,y), x,y\in\cX)$ be
a stochastic kernel. (That is, $\forall$ $x,y\in\cX$,  $\Pi (x,y)\geq 0$ and $\diy\int_{\cX}\Pi (x,y)\nu (\rd y)=1$; in
other words,
$\Pi (x,y)$ is a transition function of a Markov chain). Set $\Psi(u)=\diy\int_{\cX} \phi(x)\Pi (u,x)\nu (\rd x)$.
Then
\be\label{eq:2.3}D^{\rm{w}}_{\Psi} (f||g)\geq \Dwphi(f{\mbox{\boldmath${\Pi}$}}\,\|\,g{\mbox{\boldmath${\Pi}$}})\ee
where $\big(f{\mbox{\boldmath${\Pi}$}}\big)(x)=\diy\int_{\cX}f(u)\Pi (u,x)\nu (\rd u)$ and
$\big(g{\mbox{\boldmath${\Pi}$}}\big)(x)=\diy\int_{\cX}g(u)\Pi (u,x)\nu (\rd u)$. The equality occurs iff
$f{\mbox{\boldmath${\Pi}$}}=f$ and $g{\mbox{\boldmath${\Pi}$}}=g$.
\et

\bp
(a) The log-sum inequality yields
\be  \begin{array}{l}
 \diy\lambda_1 \phi(x) f_1(x) \log \frac{\lambda_1\phi(x) f_1(x)}{\lambda_1 g_1(x)}+\lambda_2\phi(x)f_2(x)
 \log \frac{\lambda_2\phi(x) f_2(x)}{\lambda_2 g_2(x)}\\
\qquad\qquad\diy\geq \left(\lambda_1 \phi(x) f_1(x)+\lambda_2\phi(x)f_2(x)\right)\log \frac{\lambda_1 \phi(x) f_1(x)
+\lambda_2\phi(x)f_2(x)}{\lambda_1g_1(x)+\lambda_2 g_2(x)},\;\;x\in\cX .\end{array}\ee
Integrating in $\nu (\rd x)$ yields the asserted inequality \eqref{eq:2.2}. The cases of equality emerge from the
log-sum equality cases.

(b) Again, a straightforward application of the log-sum inequality gives the result.
\ep

\bt\label{data proc.thm} Let $\BX_1^3$ be a triple of  RVs with joint PM/DF $f(\bx_1^3)$.
Let $\bx_1^3\in\cX_1^3\mapsto \phi(\bx_1^3)$ be a WF such that $X_1$ and $X_3$ are conditionally
independent given $X_2$ modulo $\phi$. (This property can be referred to as a Markov property modulo $\phi$.)

{\rm{(a)}} {\rm (Data-processing inequality for conditional WE.)} Assume inequality \eqref{eq:2.4}
(which is \eqref{eq:1.19} with $X_1$ and $X_2$ swapped):
\be \label{eq:2.4} \int_{\cX_1^3}\phi(\bx_1^3)\left[f(\bx_1^3)-f_1(x_1)\prod_{i=2,3}f_{i|1}(x_i|x_1)\right]\nu_1^3(\rd \bx_1^3)\geq 0.\ee

Then the conditional WEs satisfy property \eqref{eq:2.5}:
\be \label{eq:2.5}h^{\rm w}_{\psi_{32}}(X_3|X_2) \leq h^{\rm w}_{\psi_{31}}(X_3|X_1),\ee
with equality iff $X_2$ and $X_3$ are independent modulo $\phi$. Furthermore, assume in addition that
bound \eqref{eq:2.6} holds true
\be \label{eq:2.6} \int_{\cX_1^3}\phi(\bx_1^3)f(\bx_1^3)\Big[f_{2|13}(x_2|\bx_{13})-1\Big]\nu_1^3(\rd \bx_1^3)\leq 0\ee
(which becomes \eqref{eq:1.16} after a cyclic substitution $X_1\to X_2\to X_3\to X_1$) and suppose
$\hw_{\psi_{32}^{\,}}(X_3|X_2)=\hw_{\psi_{21}^{\,}}(X_2|X_1)$ (a stationarity-type property).
Then
\be \label{eq:2.7} h^{\rm w}_{\psi_{31}}(X_3|X_1)\leq 2 h^{\rm w}_{\psi_{32}}(X_3|X_2).\ee

{\rm{(b)}} {\rm (Data-processing inequality for mutual WE; cf. \cite{CT1}, Theorem 2.8.1.)}
Assume inequality \eqref{eq:2.8}:
\be\label{eq:2.8}\begin{array}{l}
\diy\int_{\cX_1^3}\phi (\bx_1^3)\Big[f(\bx_1^3)-f_3(x_3)\prod\limits_{i=1,2}f_{i|3}(x_i|x_3)
\Big]\nu_1^3(\rd\bx_1^3)\geq 0\end{array}\ee
(similar to \eqref{eq:1.19}, with $X_3$ and $X_2$ swapped). Then
\be\label{eq:2.9} i^{\rm w}_{\psi_{13}^{\,}}(X_1:X_3) \leq i^{\rm w}_{\psi_{12}^{\,}}(X_1:X_2).\ee
Here, equality in \eqref{eq:2.9} holds iff, modulo $\phi$, RVs $X_1$ and $X_2$ are conditionally
independent given $X_3$.
\et

\bp (a) Following the argument in Lemma \ref{lem:1.3}, we observe that
$$ \hwphi(X_3|\BX_1^2)\leq h^{\rm w}_{\psi_{31}^{\,}}(X_3|X_1).$$
On the other hand, owing to conditional independence,
\be \label{eq:2.10}\hwphi(X_3|\BX_1^2)=h^{\rm w}_{\psi_{32}^{\,}}(X_3|X_2).\ee
This yields the inequality in (\ref{eq:2.5}); for equality we need that, modulo $\phi$, RVs $X_2$
and $X_3$ are conditionally independent given $X_1$. Together with conditional independence
of $X_1$ and $X_3$ given $X_2$, it implies that for $i=1,2$, the conditional PM/DF $f_{3|i}$ does
not depend on $i$.

Next, using  Lemma \ref{lem:1.2}, we can write
\be  \diy h^{\rm w}_{\psi_{31}^{\,}}(X_3|X_1)\leq \hwphi(\BX_2^3|X_1)
:=\hwphi(X_3|\BX_1^2)+h^{\rm w}_{\psi_{21}}(X_2|X_1).\ee
Applying (\ref{eq:2.10}) yields the following assertion:
\be  \diy h^{\rm w}_{\psi_{31}^{\,}}(X_3|X_1) \leq h^{\rm w}_{\psi_{32}^{\,}}(X_3|X_2)
+h^{\rm w}_{\psi_{21}^{\,}}(X_2|X_1).\ee
Now, the assumption that $ h^{\rm w}_{\psi_{32}^{\,}}(X_3|X_2)= h^{\rm w}_{\psi_{21}^{\,}}(X_2|X_1)$
implies \eqref{eq:2.7}. The cases of equality follow from Lemmas \ref{lem:1.3} and  \ref{lem:1.2}.

(b) As before, we use Lemma \ref{lem:1.3} and Eqn \eqref{eq:2.10} (implied by conditional independence):
$$\hw_{\psi_{12}^{\,}}(X_1|X_2)=\hw_\phi (X_1|\BX_2^3)\leq \hw_{\psi_{13}^{\,}}(X_1|X_3).$$
Consequently,
$$i^{\rm w}_{\psi_{12}^{\,}}(X_1:X_2)=\hw_{\psi_1^{23}}(X_1)-\hw_{\psi_{12}^{\,}}(X_1|X_2)\geq
\hw_{\psi_1^{23}}(X_1)-\hw_{\psi_{13}^{\,}}(X_1|X_3)=i^{\rm w}_{\psi_{13}^{\,}}(X_1:X_3),$$
with the case of equality also determined from Lemma \ref{lem:1.2}.
\ep

\bt\label{concavmutual}\rm{(Cf. \cite{CT1}, Theorem 2.7.4.)}
Let $\BX_1^2$ be a pair of RVs with joint PM/DF $f(\bx_1^2)=\\
f_1(x_1)f_{2|1}(x_2|x_1)=f_2(x_2)f_{1|2}(x_1|x_2)$.
\begin{itemize}
\item {\rm{(I)}} The mutual WE
$i^{\rw}_\phi(X_1:X_2)$ is convex in $f_{2|1}(x_2|x_1)$ for fixed $f_1(X)$.
\item {\rm{(II)}} Suppose that the WF
$\vphi (x_1,x_2)$ depends only on $x_2$: $\vphi (x_1,x_2)=\vphi (x_2)$. Then
$i^{\rw}_\phi(X_1:X_2)$ is
a concave function in $f_1(X)$ for fixed $f_{2|1}(x_2|x_1)$.
\end{itemize}
\et

\bp {\rm{(I)}} For a fixed $f_1$, take two conditional PM/DFs, $f^{(1)}_{2|1}(x_2|x_1)$
and $f^{(2)}_{2|1}(x_2|x_1)$, and set
$${\wt f}_{2|1}(x_2|x_1)=\lam_1f^{(1)}_{2|1}(x_2|x_1)+\lam_2f^{(2)}_{2|1}(x_2|x_1)$$
and
$${\wt f}(\bx_1^2)= f_1(x_1){\wt f}_{2|1}(x_2|x_1)=
\lam_1f^{(1)}(\bx_1^2)+\lam_2f^{(2)}(\bx_1^2)$$
where $f^{(j)}(\bx_1^2)=f_1(x_1)f^{(j)}_{2|1}(x_2|x_1)$, $j=1,2$. Also, set:
$${\wt f}_2(x_2) =\int_{\cX_1}{\wt f}(\bx_1^2)\nu_1^2(\rd\bx_1^2)\;
\hbox{ and }\;f^{(j)}_2(x_2)
=\int_{\cX_1}f^{(j)}(\bx_1^2)\nu_1^2(\rd\bx_1^2)$$
and
$${\wt g}(\bx_1^2)=f_1(x_1){\wt f}_2(x_2),\;\hbox{and }\;g^{(j)}(\bx_1^2)=f_1(x_1)f^{(j)}_2(x_2),\;\;j=1,2.$$
Next, the mutual WE $i^\rw_\phi (X_1:X_2)$ for joint PM/DFs ${\wt f}(\bx_1^2)$
and $f^{(j)}(\bx_1^2)$ is given, respectively,  by relative WEs
$$D^\rw_\phi ({\wt f}\|{\wt g})\;\hbox{ and }\;D^\rw_\phi (f^{(j)}\|g^{(j)}),
\;j=1,2.$$
Now assertion {\rm{(I)}} follows from Theorem \ref{thm:2.2} (a).

{\rm{(II)}} Under the condition of the theorem, the reduced WF does not depend on the choice
of PM/DF $f_1$
$$\psi_2(x_2)=\int_{\cX_1}\phi(x_1,x_2)f_{1|2}(x_1|x_2)\nu_1(dx_1)=\vphi (x_2).$$
Next, write
$$\begin{array}{cl}i^\rw_\vphi (X_1:X_2)&= h^\rw_{\psi_2}(X_2) - h^{\rw}_{\phi}(X_2|X_1) \\
\;& =h^\rw_{\phi}(X_2) - \int_{\cX_1^2} \phi(x_2)f_1(x_1)f_{2|1}(x_2|x_1)\log f_{2|1}(x_2|x_1)
\nu_1^2(\rd \bx_1^2)\\
\;& =h^\rw_{\phi}(X_2) - \int_{\cX_1}f_1(x_1)h^\rw_\vphi (X_2|X_1=x_1)\nu_1(\rd x_1)\end{array}$$
where
$$h^\rw_\vphi (X_2|X_1=x_1)=\int_{\cX_2}\vphi (x_2)f_{2|1}(x_2|x_1)\log f_{2|1}(x_2|x_1)
\nu_2(\rd x_2).$$
Owing to Theorem 2.1, for fixed WF $x_2\mapsto\phi (x_2)$ and conditional PM/DF $f_{2|1}(x_2|x_1)$, the WE
$h^\rw_{\phi}$ is concave in $f_1$. The negative term is linear in $f_1$. This completes the proof of statement
{\rm{(II)}}.
\ep

\def\rA{{\rm A}}

\bt\label{Fano ineq.}
{\rm (The weighted Fano inequality; cf. \cite{CT1}, Theorem 2.10.1, \cite{KS2}, Theorem 1.2.8.)} 

Suppose an RV $X$ takes a value $x^*\in\cX$ with probability $p^*=\bbP (X=x^*)<1$
(i.e., $p^*=f(x^*)\nu (\{x^*\})$). Given a WF $x\in\cX\mapsto\phi (x)$, assume that
\be\label{eq;2.13}\diy\int_{\cX\setminus\{x^*\}}\phi (x)\left[f(x)-\frac{1-p^*}{\nu (\cX\setminus\{x^*\})}\right]\nu
(\rd x)\geq 0.\ee
Then
\be\label{eq;2.14} \hwphi (X)\leq -\phi(x^*)p^* \log\, p^*+\phi_*\log\,\left(\frac{\nu (\cX\setminus\{x^*\})}{1-p^*}\right).\ee
Here $\phi_*=\int_{\cX}\phi(x)\nu(\rd x) - \phi(x^*)p^*$.

The equality in \eqref{eq;2.14} is achieved iff $\phi (x)\left[f(x)-\diy\frac{1-p^*}{\nu (\cX\setminus\{x^*\})}\right]=0$,
for $f$-almost all $x\in\cX\setminus\{x^*\}$, i.e., iff RV $X$ is (conditionally) uniform on $\cX\setminus\{x^*\}$ modulo $\phi$.

\et

\bp
We write
\be\label{eq;2.16}\begin{array}{cl}\hwphi (X)&\diy =-\phi (x^*)p^*\log\,p^*-\int_{\cX\setminus\{x^*\}}\phi (x)f(x)\log\,f(x)\nu (\rd x)\\
\;&=\diy -\phi (x^*)p^*\log\,p^*-\log\,(1-p^*)\int_{\cX\setminus\{x^*\}}\phi (x)f(x)\nu (\rd x)\\
\;&\qquad\diy -(1-p^*)\int_{\cX\setminus\{x^*\}}\phi (x)\frac{f(x)}{1-p^*}\log\,\frac{f(x)}{1-p^*}\nu (\rd x).
\end{array}\ee
Theorem \ref{thm:1.2}, with $\beta=\diy \frac{1}{\nu (\cX\setminus\{x^*\})}$, yields that the last line in Eqn \eqref{eq;2.16} is upper-bounded by
$\phi_*\log\; \nu (\cX\setminus\{x^*\})$. This leads to  \eqref{eq;2.14}.
\ep
\def\eps{{\epsilon}}

\bt
{\rm (The weighted generalized Fano inequality; cf. \cite{KS2}, Theorem 1.2.11.)} Let $X_i:\Om\to\cX_i$, be a pair of RVs, $i=1,2$.
Suppose that $X_2$ takes exactly $m$ values $1,\ldots ,m$ (that is, $\cX_2=\{1,\ldots ,m\}$) while $X_1$
takes values  $1,\ldots, m$ and possibly other values (that is,  $\cX_1\supseteq\{1,\ldots ,m\}$), and
set: $\eps_j=\bbP(X_1\not=j|X_2=j)$. Let a WF\; $(x_1,x_2)\in\cX_1^2\mapsto \phi (x_1,x_2)$ be given such that
for all $j=1,\ldots ,m$,
\be\label{eq:2.17}\int_{\cX_1\setminus\{j\}}\phi (x_1,j)\left[f_{1|2}(x_1|j)-\frac{\eps_j}{\nu (\cX_1\setminus\{j\})}\right]\nu_1(\rd x_1)\geq 0.\ee
Then
\be\label{eq:2.18}
\begin{array}{ll}
\hw_\phi (X_1|X_2)\leq \\
\diy \sum\limits_{1\leq j\leq m}\bbP (X_2=j)\Big[
-\phi^*_j(0)(1-\eps_j)\log\; (1-\eps_j)+\phi^*_j(1) \log \Big(\frac{\nu_1(\cX_1\setminus\{j\})}{\eps_j}\Big)\Big].
\end{array} \ee
Here RV $X^*_j$ takes two values, say $0$ and $1$, with $\bbP (X^*=0)=1-\eps_j=1-\bbP (X^*=1)$, and
the WF $\phi^*$ has
\be\label{eq:2.19}\phi^*_j(0)=\phi (j,j)\;\hbox{ and }\;\phi^*_j(1)=\int_{\cX\setminus\{j\}}
\phi (x_1,j)f(x,j)\nu_1(\rd x_1).
\ee
\et

\bp
By definition of the conditional WE, the weighted Fano inequality, Theorem \ref{thm:1.2}
and with definitions \eqref{eq:2.19} at hand,
we obtain that
$$\begin{array}{cl}\hw_\phi (X_1|X_2)&\leq \;\;\sum\limits_j\bbP (X_2=j)
\bigg[-\phi (j,j)(1-\eps_j)\log\,(1-\eps_j)\\
\;&\qquad +\diy\int_{\cX\setminus\{j\}}\phi (x_1,j)f(x,j)\nu_1(\rd x_1)\log\;\frac{\nu_1(\cX_1
\setminus\{j\})}{\eps_j}\bigg].\end{array}$$
This yields inequality \eqref{eq:2.18}.
\ep

\section{Maximum WE properties}

In this section we establish some extremality properties for the WE; cf. \cite{CT2}, Chap. 12.

\bt
Suppose $X^*:\Om\to\cX$ is an RV with a PM/DF $f^*$ and $x\in\cX\to\phi (x)$ is a given WF.
\begin{itemize}
  \item {\rm{(I)}} Then $f^*$ (or $X^*$) is the unique maximizer, modulo $\phi$, of  the WE
$h^\rw_\phi (f)$ under the constraints
 \be\label{eq:constr}\diy\int_\cX\phi (x )\big[f(x)-f^*(x)\big]\nu (\rd x)\geq 0\;\hbox{ and }\;\\
 \int\limits_\cX\phi (x )\big[f(x)-f^*(x)\big]\log f^*(x) \nu (\rd x)\geq 0.\ee

  \item {\rm{(II)}} On the other hand, consider a constraint
\be\label{eq:constr1}\int_\cX\phi (x)f(x)\beta (x)\rd\nu (x)= c
\ee
where $x\in\cX\mapsto\beta (x)$ is a given function and $c$ a given constant neither of
which is assumed non-negative. Suppose that
$f^*(x)=\diy\frac{1}{Z}\exp [-b\beta (x)]$ is a (Gibbsian-type)  PM/DF such that
$$\int_\cX\phi (x)f^*(x)\rd\nu (x)= 1\;\hbox{ and }\;
\int_\cX\phi (x)f^*(x)\beta (x)\rd\nu (x)= c.$$
Here $b$ is a constant (an analog of  inverse temperature) and
$Z=\int_\cX\exp [-b\beta (x)]\rd\nu (x)\in (0,\infty )$
is the normalizing denominator (an analog of a partition function). Introduce the second
constraint:
\be\label{eq:constr2}(\log Z)\int_\cX\phi (x)[f^*(x)-f(x)]\rd\nu (x)\geq 0.\ee
Then, under \eqref{eq:constr1} and  \eqref{eq:constr2}, the WE  $h^{\rm w}_\phi (f)$ is
maximized at $f=f^*$. As above, it is a unique maximizer, modulo $\phi$.
\end{itemize}
\et

\bp
(I) Using definition \eqref{eq:1.02} and Theorem \ref{thm:1.1}, we obtain
\be\label{eq:Gibbsappl}
0 \geq -\Dwphi(f\|f^*)=\hwphi(f)+\int\limits_\cX \phi( x) f( x) \log f^*( x) \nu (\rd x).
\ee
Under our constraint \eqref{eq:constr} it yields
\be  \hwphi(f)\leq -\int\limits_\cX \phi( x)f^*(x) \log f^*(x) \nu (\rd x)=\hwphi(f^*).\ee
The uniqueness of the maximizer follows from the uniqueness case for equality in the
weighted Gibbs inequality.

(II) Again use \eqref{eq:Gibbsappl}:
$$\begin{array}{cl}h^{\rm w}_\phi (f)&\diy\leq -\int_\cX\phi (x)f(x)\Big[-\log\,Z-b\beta (x)\Big]\rd\nu (x)\\
\;&\diy=(\log\,Z)\int_\cX\phi (x)f(x)\rd\nu (x)+b\int_\cX\phi (x)f(x)\beta (x)\rd\nu (x)\\
\;&\diy \leq (\log\,Z)\int_\cX\phi (x)f^*(x)\rd\nu (x)+b\int_\cX\phi (x)f^*(x)\beta (x)\rd\nu (x)=h^{\rm w}_\phi (f^*).
\ena$$
\ep

Note that when $Z\geq 1$, the factor $\log\,Z$ can be omitted from \eqref{eq:constr2}; otherwise
$\log\,Z$ can be replaced by $-1$.

\def\BC{{\mathbf C}} \def\bbR{{\mathbb R}} \def\bmu{{\mbox{\boldmath${\mu}$}}}
 \def\bPhi{{\mbox{\boldmath${\Phi}$}}}   \def\bPsi{{\mbox{\boldmath${\Psi}$}}}
 \def\bbZ{{\mathbb Z}}
 \def\rT{{\rm T}}

\bex\label{ex:mvnormal}
Consider a random vector $\BX=\BX_1^d:\Om\to\bbR^d$ with PDF $f$ (relative to the $d$-dimensional
Lebesgue measure), mean vector $\bf{0}$ and covariance matrix $\BC=(C_{ij})$ with with $C_{ij}=\bbE[X_iX_j]$, $1\leq i,j\leq d$. Let $f^{\rm{No}}_\BC$ be the normal PDF with the same ${\mbox{\boldmath${\mu}$}}$ and $\BC$.
Let $\bx=\bx_1^d\in\bbR^d\mapsto\phi (\bx)$ be a given WF which is positive on an
open domain in $\bbR^d$. Introduce $d\times d$ matrices $\bPhi=(\Phi_{ij})$,
$\bPhi^{\rm{No}}_\BC=(\Phi^{\rm{No}}_{ij})$ and
$\bx^{\rT}\bx$,
where $(\bx^{\rT}\bx)_{ij}=x_i x_j$ and
$$\Phi =\int_{\bbR^d}\phi (\bx)f(\bx)\bx^{\rT}\bx \rd \bx,\;\;
\Phi^{\rm{No}}_\BC=\int_{\bbR^d}\phi (\bx)f^{\rm{No}}_\BC(\bx)\bx^{\rT}\bx \rd\bx.$$
Suppose that
\be\label{3.3}\begin{array}{l}
\diy\int_{\bbR^d}\phi (\bx)\Big[f(\bx)-f^{\rm{No}}_\BC (\bx)\Big]\rd\bx\geq 0
\;\;\hbox{ and }\\
\diy\log \left[ (2\pi)^d ({\rm{det}}\,\BC)\right]\int_{\bbR^d}\phi (\bx)\Big[f(\bx)-f^{\rm{No}}_\BC
 (\bx)\Big]\rd\bx+{\rm{tr}}\,\Big[\BC^{-1}\left(\bPhi-\bPhi^{\rm{No}}_\BC
 \right)\Big]\leq 0.
\end{array}\ee
Then
\be  \hwphi(f)\leq \hwphi (f^{\rm{No}}_\BC)=\frac{1}{2} \log \left[ (2\pi)^d ({\rm{det}}\,\BC)\right]\int\limits_{\bbR^d}\phi (\bx)f^{\rm{No}}_\BC(\bx)\rd\bx+\frac{\log\,e}{2}{\rm{tr}}\,\BC^{-1}\bPhi^{\rm{No}}_\BC, \ee
with equality iff $f=f^{\rm{No}}_\BC$ modulo $\phi$.
\eex

\bp
Using the same idea as before, write
\be
0 \geq -\Dwphi(f\|f^{\rm{No}}_\BC)=
\hwphi(f) -\frac{1}{2} \log \left[ (2\pi)^d ({\rm{det}}\,\BC)\right] \int\limits_{\bbR^d}\phi (\bx )f(\bx)\rd\bx-\frac{\log\,e}{2}{\rm{tr}}\,\BC^{-1} \bPhi,\ee
Equivalently,
$$ \hwphi(f)\leq \frac{1}{2} \log \left[ (2\pi)^d ({\rm{det}}\,\BC)\right]\int\limits_{\bbR^d}\phi (\bx)f(\bx)\rd\bx+\frac{\log\,e}{2}{\rm{tr}}\,\BC^{-1}\bPhi$$
which leads directly to the result.
\ep

To further illustrate the above methodology, we provide some more examples, omitting the proofs.

\bex Let $f^{\rm{Exp}}$ denote an exponential PDF on $\bbR_+=(0,\infty )$ (relative to
the Lebesgue measure $\rd x$) with mean $\lam^{-1}$. Suppose a PDF $f$ on $\bbR_+$ satisfies
the constraints
\be\begin{array}{l}
\diy\int\limits_{\bbR_+}\phi (x )\big[f(x)-f^{\rm{Exp}}(x)\big]\rd x\geq 0\;\;\hbox{ and }\\
\diy\big(\log\,\lam\big)\int\limits_{\bbR_+}\phi (x )\big[f(x)-f^{\rm{Exp}}(x)\big]\rd x-
\lam\;\int\limits_{\bbR_+}x\;\phi (x)\big[f(x)-f^{\rm{Exp}}(x)\big]\rd x\geq 0.
\end{array}\ee
where $x\in\bbR_+\mapsto\phi (x)$ is a given WF positive on an open interval. Then
$$ \hwphi(f)\leq \hwphi(f^{\rm{Exp}})=-\big(\lam\log\,\lam\big)\int_{\bbR_+}\phi (x )e^{-\lam x}\rd x+\lam^2\;
\int_{\bbR_+}x
\phi (x)e^{-\lam x}\rd x,$$
and $f^{\rm{Exp}}$ is a unique maximizer modulo $\phi$.
\eex


\bex Take $\cX=\bbZ_+=\{0,1,\ldots\}$ and let $\nu$ be the counting measure: $\nu (i)=1$ $\forall$
$i\in\bbZ_+$. Then, for a RV $X$ with PMF $f(i)$ we have $f(i)=\bbP (X=i)$. Fix a WF\ $i\in\bbZ_
+\mapsto \phi (i)$.

{\rm{(a)}} Let $f^{\rm{Ge}}$ be a geometric PMF: $f^{\rm{Ge}}(x)=(1-p)^xp$, $x\in\bbZ_+$. Then for any
PMF $f(x)$,
$i\in\bbZ_+$, satisfying the constraints
\be\begin{array}{l}
\diy\sum\limits_{i\in\bbZ_+}\phi(i)\big[f(i)-f^{\rm{Ge}}(i)\big]\geq 0\;\;\hbox{ and }\\
\log\;p\;\diy\sum\limits_{i\in\bbZ_+}\phi(i)\big[f(i)-f^{\rm{Ge}}(i)\big]+
\log (1-p)\;\diy\sum\limits_{i\in\bbZ_+} i\phi(i)\big[f(i)-f^{\rm{Ge}}(i)\big] \geq 0.
\end{array}\ee
we have $\hw_\phi(f)\leq \hw_\phi (f^{\rm{Ge}})$, with equality iff  $f=f^{\rm{Ge}}$ modulo $\phi$.

{\rm{(b)}} Let $f^{\rm{Po}}$ be a Poissonian PMF: $f^{\rm{Po}}(k)=\diy\frac{e^{-\lam}\lam^k}{k!}$, $k\in\bbZ_+$.
Then for any PMF $f(k)$,
$k\in\bbZ_+$, satisfying the constraints
\be\begin{array}{l}
\diy\sum\limits_{k\in\bbZ_+}\phi(k)\big[f(k)-f^{\rm{Po}}(k)\big]\geq 0\;\;\hbox{ and }\\
\log\; \lambda \sum\limits_{k\in\bbZ_+}k\phi(k)\big[f(k)-f^{\rm{Po}}(k)\big]\\
\quad-\lambda\; \sum\limits_{k\in\bbZ_+}\phi(k)\big[f(k)-f^{\rm{Po}}(k)\big]
- \sum\limits_{k\in\bbZ_+} (\log\,k!)\phi(k)\big[f(k)-f^{\rm{Po}}(k)\big] \geq 0.
\end{array}\ee
we have $\hw_\phi(f)\leq \hw_\phi (f^{\rm{Po}})$, with equality iff  $f=f^{\rm{Po}}$ modulo $\phi$.
\eex

Theorem \ref{KyFan} below offers an extension of the Ky Fan inequality that $\log \rm{det}\;\BC$ is a concave function of a positive definite $d\times d$ matrix $\BC$. Cf. \cite{KF1,KF2,KF3,Mo}. We follow the method proposed by Cover-Dembo-Thomas. As before,
$f^{\rm{No}}_{\BC}$ denotes the normal PDF with zero
mean and covariance matrix $\BC$.

\bt\label{KyFan}
{\rm (The weighted Ky Fan inequality; cf. \cite{CT1}, Theorem 17.9.1, \cite{CT2}, Theorem 1, \cite{DCT}, Theorem 8, \cite{KS2}, Worked Example 1.5.9.)} Assume that $\bx_1^d \in \bbR^d \mapsto \phi (\bx_1^d)\geq 0$ is a given WF positive on an open domain. Suppose that,
for $\lam_1,\lam_2\in [0,1]$ with $\lam_1+\lam_2=1$ and positive-definite $\BC_1$, $\BC_2$, with
 $\BC=\lam_1\BC_1+\lam_2\BC_2$,
\be\label{eq:KyFanCond}
\diy\int_{\bbR^d}\phi (\bx)\Big[\lam_1f^{\rm{No}}_{\BC_1}(\bx)+\lam_2f^{\rm{No}}_{\BC_2}(\bx)-
f^{\rm{No}}_\BC(\bx)\Big]\rd\bx\geq 0,\;\hbox{ and}\ee
\be\label{eq:KyFanCond1} \log \left[ (2\pi)^d ({\rm{det}}\,\BC)\right]\int_{\bbR^d}\phi (\bx)\Big[\lam_1f^{\rm{No}}_{\BC_1}(\bx)+\lam_2f^{\rm{No}}_{\BC_2}(\bx)-
f^{\rm{No}}_\BC(\bx)\Big]\rd\bx  +\frac{\log e}{2}{\rm{tr}}\Big[\BC^{-1}\bPsi\Big]\leq 0,\ee \be \hbox{where}\;\;\;\; \diy\bPsi
=\int_{\bbR^d}\phi (\bx)\Big[\lam_1f^{\rm{No}}_{\BC_1}(\bx)+\lam_2f^{\rm{No}}_{\BC_2}(\bx)-
f^{\rm{No}}_\BC(\bx)\Big](\bx-\bmu)^{\rT}(\bx-\bmu)\rd \bx.\ee

Then, with $\sigma_\phi (\BC)=\hwphi (f^{\rm{No}}_{\BC})$, $\sigma_\phi (\BC_1)=\hwphi (f^{\rm{No}}_{\BC_1})$ and $\sigma_\phi (\BC_2)=\hwphi (f^{\rm{No}}_{\BC_2})$
\be\label{eq:KyFan}\sigma_\phi (\BC)-\lam_1\sigma_\phi (\BC_1)-\lam_2\sigma_\phi (\BC_2)\geq 0;
\ee
equality iff $\;\lam_1\lam_2=0\;$ or $\;\BC_1=\BC_2$.
\et

\bp
Take values $\lambda_1,\lambda_2\in[0,1]$, such that $\lambda_1+\lambda_2=1$. Let $\BC_1$ and
$\BC_2$ be two positive definite $d\times d$ matrices. Let $\BX_1$ and $\BX_2$
be two multivariate normal vectors, with PDFs $f_k\sim {\rm N}(0,\BC_k)$, $k=1,2$. Set $\BZ=\BX_{\Theta}$,
where the RV
$\Theta$, takes two values, $\theta =1$ and $\theta =2$ with probability $\lambda_1$ and $\lambda_2$
respectively, and is independent of $\BX_1$ and $\BX_2$. Then vector $\BZ$
has covariance $\BC=\lambda_1\BC_1+\lambda_2\BC_2$. Also set:
\be\label{eq:alpha}\alpha (\BC)=\int\limits_{\bbR^d}\phi (\bx )f^{\rm{No}}_{\BC}(\bx )\rd\bx .\ee

Let $\bx=(x_1,\ldots ,x_d)\in\bbR^d\mapsto\phi (\bx)$ be a given WF and set $\wt\phi (\bx,\tha )=
\phi (\bx)$.
 Following the same arguments as in the proof of Theorem \ref{thm2-1}, $\hw_{\wt\phi}(\BZ|\Theta)\leq
 \hwphi (\BZ)$. It is plain that
\be \begin{array}{cl} \hw_{\wt\phi}(\BZ|\Theta)&=\lambda_1\hwphi(X_1)+\lambda_2\hwphi(X_2)\\
\;&\diy=\sum\limits_{k=1,2}
\lam_k\left\{\frac{1}{2}\log \left[ (2\pi)^d ({\rm{det}}\,\BC_k)\right]\int\limits_{\bbR^d}\phi (\bx)f^{\rm{No}}_{\BC_k}(\bx)\rd
\bx+\frac{\log\,e}{2}
{\rm{tr}}\,\BC^{-1}_k\bPhi^{(k)}\right\}\end{array}\ee
where
$$\bPhi^{(k)}=\int\limits_{\bbR^d}\bx^{\rT}\bx\;\phi (\bx)f^{\rm{No}}_{\BC_k}(\bx)\rd\bx,
\;\;k=1,2,$$
and $(\bx^{\rT}\bx)_{ij}=x_ix_j$.
According to Example \ref{ex:mvnormal}, we have
\be  \hwphi (\BZ)\leq \frac{1}{2}\Big\{ \log \left[ (2\pi)^d ({\rm{det}}\,\BC)\right]\Big\}
\alpha (\BC)+\frac{\log\,e}{2}{\rm{tr}}\,\BC^{-1}\bPhi , \ee
where
\be\label{eq:bPhi}\bPhi=\int\limits_{\bbR^d}\bx^{\rT}\bx\;\phi (\bx)f^{\rm{No}}_{\BC}(\bx)\rd\bx.
\ee
The inequality \eqref{eq:KyFan} then follows. The cases of equality are covered by Theorem \ref{thm2-1}.
\ep

The following lemma is an immediate extension of Lemma \ref{lem:1.1}.

\bl\label{Multi-subadditivity}
Let $\BX_1^n=(X_1,\ldots ,X_n)$ be a random vector, with components
$X_i:\Om\to\cX_i$, $1\leq i\leq n$,
and the joint PM/DF $f$. Extending the notation used in Sect $1$, set:
$$\bx_1^n=(x_1,\ldots, x_n)\in\cX_1^n:=\operatornamewithlimits{\times}\limits_{1\leq i\leq n}\cX_i\
\hbox{ and }\;
\nu_1^n(\rd\bx_1^n)=\prod\limits_{1\leq i\leq n}\rd\nu_i(\rd x_i),$$
and more generally,
$$\bx_k^l=(x_k,\ldots, x_l)
\in\cX_k^l:=\operatornamewithlimits{\times}\limits_{k\leq i\leq l}\cX_i\hbox{ and }\;
\nu_k^l(\rd\bx_k^l)=\prod\limits_{k\leq i\leq l}\rd\nu_i(\rd x_i),\;\;1\leq k\leq l.$$
Next, introduce
$$\begin{array}{r}\diy f_i(x_i)=\int_{\cX_1^{i-1}\times\cX_{i+1}^n}f(\bx_1^{i-1},x_i,\bx_{i+1}^n)\nu_1^{i-1}
(\rd\bx_1^{i-1})
\nu_{i+1}^n(\rd\bx_{i+1}^n)\qquad{}\hbox{(the marginal PM/DF for RV $X_i$)},\end{array}$$
and
$$f_{\;|i\,}(\bx_1^n|x_i)=\frac{f(\bx_1^n)}{f_i(x_i)}\qquad{}\hbox{(the conditional PM/DF given that
$X_i=x_i$).}$$
Given a WF $\bx_1^n\in\cX_1^n\mapsto\phi (\bx_1^n)$, suppose that
\be \int_{\cX_1^n}\phi(\bx_1^n)\left[f(\bx_1^n)-\prod\limits_{i=1}^n f_i(x_i)\right]
\nu_1^n (\rd\bx_1^n)\geq 0.\ee
Then
\be \label{subaditivity vector} \hwphi(\BX_1^n)\leq \sum\limits_{i=1}^n \hw_{\psi_i}(\BX_i),\ee
where
$$\psi_i(x_i)=\diy\int_{\cX_1^{i-1}\times\cX_{i+1}^n}\phi(\bx_1^n)f_{\;|i\,}(\bx_1^n|x_i)
\nu_1^{i-1}(\rd\bx_1^{i-1})\nu_{i+1}^n(\rd\bx_{i+1}^n).$$
Here, equality in \eqref{subaditivity vector} holds iff, modulo $\phi$, components $X_1,\dots,X_n$ are independent.
\el

\bt\label{Hadamard}
{\rm (The weighted Hadamard inequality; cf. \cite{CT1}, Theorem 17.9.2, \cite{CT2}, Theorem 3, \cite{DCT}, Theorem 26, \cite{KS2}, Worked Example 1.5.10).} Let  $\BC=(C_{ij})$ be a positive definite $d\times d$ matrix
and $f^{\rm{No}}_{\BC}$ the normal PDF with zero mean and covariance matrix $\BC$. Given a WF
function $\bx_1^d=(x_1,\ldots ,x_d)\in\bbR^d\mapsto\phi (\bx_1^d)$, positive on an open domain
in $\bbR^d$, introduce quantity $\alpha =\alpha (\BC )$ by
\eqref{eq:alpha} and matrix $\bPhi =(\Phi_{ij})$ by
\eqref{eq:bPhi}.  Let $f^{\rm{No}}_i$ stand for the ${\rm N}(0,C_{ii})$-PDF
(the marginal PDF of the $i$-th component).
Then under condition
\be\int_{\bbR^d}\phi (\bx_1^d)\Big[f^{\rm{No}}_{\BC}(\bx_1^d)-\prod\limits_{i=1}^d f^{\rm{No}}_i(x_i)\Big]\rd\bx_1^d\geq 0,\ee
we have:
\be\label{eq:Hadamard} \diy\alpha \log \prod\limits_{i} (2\pi C_{ii})+(\log\,e)\sum\limits_i C_{ii}^{-1}\Phi_{ii}-
\alpha \log \; \Big[(2\pi)^d({\rm{det}}\,\BC)\Big]-(\log\,e){\rm{tr}}\,\BC^{-1}\bPhi \geq 0, \ee
with equality iff $\BC$ is diagonal.
\et

\bp
If $X_1,\dots,X_d \sim \mathrm{N}(0,\BC)$, then in Lemma \ref{Multi-subadditivity}, by following (\ref{subaditivity vector}) we can write
\be  \begin{array}{l} \diy \frac{1}{2} \log \left[ (2\pi)^d ({\rm{det}}\,\BC)\right]
\int\limits_{\bbR^d}\phi (\bx_1^d)f(\bx_1^d)\rd\bx_1^d+\frac{\log\,e}{2}{\rm{tr}}\,\BC^{-1}\bPhi\\
\qquad\qquad\diy \leq\frac{1}{2}\, \sum\limits_{i=1}^d\;\left[\log \left(2\pi C_{ii}\right)
\int\limits_{\bbR}\psi_i (x )f^{\rm{No}}_i(x )\rd x+(\log\,e)\,C_{ii}^{-1}\Psi_{ii}\right].\end{array}\ee
Here
$$\psi_i(x_i)=\int_{\bbR^{d-1}}\phi (\bx_1^d)f^{\rm{No}}_{\;|i\,}(\bx_1^d|x_i)\prod_{j:j\neq i}\rd x_j,\;\;\Psi_{ii}=\int\limits_{\bbR^d} x_i^2 \psi_i (x_i)f^{\rm{No}}_i(x_i)\rd x_i =\Phi_{ii}$$
and
$$f^{\rm{No}}_{\;|i\,}(\bx_1^d|x_i)=\frac{f^{\rm{No}}_{\BC}(\bx_1^d)}{f^{\rm{No}}_i(x_i)}\;\;\hbox{(the conditional PDF)}.$$
With
$$\alpha=\int\limits_{\bbR}\psi_i (x_i)f^{\rm{No}}_i(x_i)\rd x_i=\int\limits_{\bbR^d}\phi (\bx_1^d)f^{\rm{No}}_{\BC}(\bx_1^d)\rd\bx_1^d,$$
the bound \eqref{eq:Hadamard} follows.
\ep

\br As above, maximizing the left-hand side in \eqref{eq:Hadamard} would give a bound between
${\rm{det}}\,\BC$ and the product $\prod\limits_{i=1}^dC_{ii}$.
\er

\section{Weighted Fisher information and related inequalities}

\def\utheta{{\underline\theta}}
In this section we introduce a weighted version of Fisher information matrix and establish
some straightforward facts. The bulk of these properties is derived by following Ref. \cite{Z}.

\bd
Let $\BX=(X_1, \ldots ,X_n)$ be a random $1\times n$ vector with probability density function (PDF)
$f_\utheta (\bx )=f_\BX(\bx ;\utheta)$, $\bx =(x_1, \ldots ,x_n)\in\bbR^n$, where
$\utheta =(\theta_1,\ldots , \theta_m)\in\bbR^m$ is a parameter vector. Suppose that
dependence $\utheta\mapsto f_\utheta$ is {\rm C}$^1$.
The $m\times m$ weighted Fisher information matrix $\tJ^\rw_\phi(\BX ;\utheta )$, with
a given WF $\bx\in\bbR^n\mapsto\phi (\bx )\geq 0$, is defined by
\be\label{eq:WtFI} \tJ^{\rw}_\phi (\BX;\utheta)=\bbE\Big[\phi (\BX)\BS (\BX,\utheta )^\rT \BS (\BX,\utheta )\Big]
=\int\frac{\phi(\bx)}{f_{\utheta}(\bx)}\bigg(\frac{\partial f_{\utheta}(\bx)}{\partial
\utheta}\bigg)^\rT \frac{\partial f_{\utheta}(\bx)}{\partial \utheta}
{\mathbf 1}\left(f_\utheta (\bx )>0\right)\rd \bx ,\ee
assuming the integrals are absolutely convergent.
Here and below, $\diy\frac{\partial}{\partial \utheta}$ stands for the $1\times m$ gradient in $\utheta$ and
$\BS (\BX,\utheta )={\mathbf 1}(f_\utheta (\bx )>0)\diy\frac{\partial}{\partial\utheta}\log f_\utheta (\bx)$ denotes the {\it score vector}.

When $\phi (\bx )\equiv 1$, $\tJ^\rw_\phi (\BX;\utheta )=\tJ(\BX;\utheta )$, the standard Fisher
information matrix, cf. {\rm{\cite{DCT}, \cite{CT2}, \cite{KS2}}}.
\ed


\bd
Let $(\BX,\BY)$ be a pair of RVs with a joint PDF $f_\utheta (\bx ,\by)=f_{\BX ,\BY}(\bx,\by ;\utheta)$
and conditional PDF $f_\utheta (\by |\bx )=f_{\BY|\BX}(\by |\bx ;\utheta):=\diy\frac{f_{\BX ,\BY}(\bx,\by ;\utheta)}{
f_\BX(\bx ;\utheta)}$. Given a joint WF  $(\bx ,\by )\in\bbR^n\times\bbR^n\mapsto \phi (\bx ,\by)\geq 0$, we
set:
\be\label{eq:jnWtFI}\begin{array}{cl}\tJ^{\rw}_\phi (\BX,\BY;\utheta)
&\diy=\bbE\left[\phi (\BX ,\BY)\left(\frac{\partial\log\,f_\utheta (\BX,\BY)}{\partial\utheta}\right)^\rT
\frac{\partial\log\,f_\utheta (\BX,\BY)}{\partial\utheta}
{\mathbf 1}\left(f_\utheta (\BX ,\BY )>0\right)\right]\\
\;&\diy =\int\frac{\phi(\bx ,\by)}{f_{\utheta}(\bx ,\by)}
\bigg(\frac{\partial f_{\utheta}(\bx ,\by)}{\partial \utheta}\bigg)^\rT
\frac{\partial f_{\utheta}(\bx ,\by)}{\partial \utheta}
{\mathbf 1}\left(f_\utheta (\bx ,\by )>0\right)\rd \bx\rd\by\end{array}\ee
and
\be\label{eq:condWFI} \begin{array}{l}\diy\tJ^{\rw}_\phi (\BY|\BX;\utheta )=\bbE\left[\phi (\BX,\BY)
\bigg(\frac{\partial \log f_{\utheta}(\BY|\BX)}{\partial \utheta}\bigg)^\rT
\frac{\partial \log f_{\utheta}(\BY|\BX)}{\partial \utheta}{\mathbf 1}\left(f_\utheta (\BY|\BX )
>0\right)\right]\\
\qquad\qquad =\diy\int\frac{\phi (\bx,\by)f_ \utheta (\bx,\by )}{f_{\utheta}(\by|\bx )^2}\bigg(\frac{\partial f_{\utheta}(\by|\bx )}{\partial \utheta}\bigg)^\rT
\frac{\partial f_{\utheta}(\by|\bx )}{\partial \utheta}
{\mathbf 1}\left(f_\utheta (\by|\bx )>0\right)\rd\bx\rd\by .\end{array}\ee

Next, consider an $m\times m$ matrix
 $\tS_{\utheta}=\tS_{\utheta} (f_{\BX ,\BY})$ and a $1\times m$ vector $\BB_{\utheta}=
 \BB_{\utheta} (\bx,f_{\BY|\BX})$:
\be\label{notation:1.1}
\BB_{\utheta}=\bbE_{\BY|\BX=\bx}\bigg[\phi(\bx,\BY)\frac{\partial\log f_{\utheta}(\BY|\bx)}{\partial\utheta}
\bigg]=\int \frac{\phi(\bx,\by)}{f_\utheta (\by |\bx)} \frac{\partial f_{\utheta}(\by|\bx)}{\partial\utheta}{\mathbf 1}\left(f_\utheta (\by|\bx )>0\right)\rd\by,
\ee
\be\label{tf} \tS_{\utheta}=\bbE\left \{\left[\bigg(\frac{\partial\log f_{\utheta}(\BX)}{\partial \utheta}
\bigg)^\rT\BB_{\utheta}(\BX)
+\BB_{\utheta}(\BX)^\rT\frac{\partial\log f_{\utheta}(\BX)}{\partial \utheta}\right]{\mathbf 1}\left(f_\utheta (\BX)>0\right)\right \}.\ee
When $\phi (\bx, \by)$ depends only on $\bx$ and under standard regularity assumptions, vector $\BB_{\utheta}$ vanishes (and so does matrix $\tS_{\utheta}$):
$$\diy \BB_{\utheta}= \phi(\bx)\int \frac{\partial f_{\utheta}(\by|\bx)}{\partial\utheta}\,\rd\by = \frac{\partial }{\partial \utheta} \int f_{\utheta}(\by|\bx) \rd\by = 0.$$
\ed

For the sake of brevity, in formulas that follow we routinely omit indicators of positivity of PDFs involved: their presence can be easily derived from the local context.

\bl\label{lem:4.1} {\rm{(The chain rule: cf. \cite{Z}, Lemma 1.)}} Given a pair
$(\BX,\BY)$ of random vectors and a joint WF $(\bx,\by)\mapsto \phi (\bx,\by)$, set:
\be\label{eq:psiX} \psi (\bx )= \psi_\BX (\bx )=\int \phi (\bx,\by) f_ \utheta (\by|\bx)\,\rd\by=\bbE_{\BY|\BX=\bx}\phi(\bx,\BY).\ee
Then
\be\label{Chain.rule} \tJ^\rw_\phi(\BX,\BY; \utheta)=\tJ^{\rm w}_{\psi}(\BX; \utheta)+\tJ^\rw_\phi(\BY|\BX; \utheta)
+\tS_{\utheta}.\ee
\el

\bp For simplicity, assume that $\utheta$ is scalar: $\utheta = \theta$; a generalization to a vector case is straightforward.
Therefore, we have
\be\label{eq:4.1} \tJ^{\rm w}_{\phi}(\BX,\BY; \theta)=\bbE\bigg[\phi(\BX,\BY)\bigg(\frac{\partial\log
f_{ \theta}(\BX,\BY)}{\partial  \theta}\bigg)^2\bigg].\ee
Furthermore, we know
$$\log f_{ \theta}(\bx,\by)=\log f_{ \theta}(\bx)+\log f_{ \theta}(\by|\bx)$$
Using (\ref{eq:4.1}) yields:
\be  \begin{array}{l}\tJ^{\rm w}_{\phi}(\BX,\BY; \theta)=\diy \bbE\bigg[\phi (\BX,\BY)\bigg(\frac{\partial
\log f_{ \theta}(\BX)}{\partial  \theta}\bigg)^2\bigg]\\
\quad +\diy \bbE\bigg[\phi(\BX,\BY)\bigg(\frac{\partial \log f_{ \theta}(\BY|\BX)}{\partial  \theta}\bigg)^2\bigg]+2\;\diy
\bbE\bigg[\phi(\BX,\BY)\bigg(\frac{\partial \log f_{ \theta}(\BX)}{\partial  \theta}\bigg)\bigg(\frac{\partial
\log f_{ \theta}(\BY|\BX)}{\partial  \theta}\bigg)\bigg].\end{array}\ee
We also can write

\be  \begin{array}{l} \diy\bbE\bigg[\phi(\BX,\BY)\bigg(\frac{\partial \log f_{ \theta}(\BX)}{\partial  \theta}\bigg)
\bigg(\frac{\partial \log f_{ \theta}(\BY|\BX)}{\partial  \theta}\bigg)\bigg]\\
\qquad\qquad\qquad = \diy\bbE\Bigg\{\frac{\partial\log f_{ \theta}(X)}{\partial  \theta}\bbE\bigg[\phi(\BX,\BY)\bigg(\frac{\partial
\log f_ \theta(\BY|\BX)}{\partial  \theta}\bigg)\Big|\BX\bigg]\Bigg\}. \end{array}\ee
This cancels the last term in (\ref{Chain.rule}) when applying inner expectation in the RHS of (\ref{Chain.rule}).
\ep

\def\BA{{\mathbf A}} \def\BB{{\mathbf B}}

Throughout the paper, an inequality $\tA \leq\tB$ between matrices $\tA$ and $\tB$ means that $\tB-\tA$ is a positive-definite matrix.

\bl\label{lem:4.2} {\rm{(Data-refinement inequality: cf. \cite{Z}, Lemma 2.)}} For a given joint
WF $(\bx ,\by )\mapsto \phi (\bx ,\by )$,
\be\label{eq:DRIneq}\tJ^\rw_\phi(\BX,\BY;\utheta)\geq \tJ^{\rm w}_{\psi}(\BX;\utheta)+\tS_{\utheta},\ee
with equality if $\BX$ is a sufficient statistic for $ \utheta$. Here WF $\psi =\psi_\BX$ is defined as in \eqref{eq:psiX}.
\el

\bp Bound \eqref{eq:DRIneq} follows from Lemma \ref{lem:4.1} using the non-negativity of matrix
$$\tJ^\rw_\phi(\BY|\BX=\bx ;\utheta) =\diy\int f_ \utheta (\by|\bx )\phi (\bx,\by)
\bigg(\frac{\partial \log f_{\utheta}(\by|\bx )}{\partial \utheta}\bigg)^\rT
\frac{\partial \log f_{\utheta}(\by|\bx )}{\partial \utheta}\rd\by.$$
Equality holds when $\tJ^\rw_\phi(\BY|\BX=\bx ;\utheta)=0$ which leads to the statement.
\ep

\bl\label{lem:4.3}{\rm{(Data-processing inequality: cf. \cite{Z}, Lemma 3.)}} For
a given joint WF $(\bx ,\by )\mapsto \phi (\bx ,\by )$ and
a function $\bx\mapsto g(\bx)$, set
\be{\varrho}_g(\bx)=\phi(\bx,g(\bx))\;\;\; \hbox{and}\;\;\; \rho_g(\bx)=\phi(\bx,g(\bx))f_ \utheta (\bx|g(\bx)).\ee
Then we have
\be \tJ^{\rm w}_{\varrho_g}(\BX;\utheta)\geq \tJ^{\rm w}_{\rho_g}(g(\BX);\utheta).\ee
The equality holds iff function $g(\BX)$ is invertible.
\el

\bp We make use Lemma \ref{lem:4.2}. Let $\BY=g(\BX )$, then $\tS_{\utheta}={\tt 0}$. This yields
\be\label{eq:4.2} \tJ^{\rm w}_{\rho_g}(g(\BX );\utheta)\leq \tJ^\rw_\phi(\BX,g(\BX);\utheta).\ee
Note that the equality holds true if $\tJ^\rw_\phi(\BX |g(\BX);\utheta )=0$, that is $g(\BX)$ is a
sufficient statistics for
$\utheta$. Now use the chain rule, Lemma \ref{lem:4.1}, where $\tJ^\rw_\phi(g(\BX) |\BX;\utheta )=0$. Hence,
\be \label{eq:4.3}\tJ^\rw_\phi(\BX ,g(\BX);\utheta)=\tJ^{\rm w}_{\varrho_g}(\BX;\utheta).\ee
The assertions (\ref{eq:4.2}) and (\ref{eq:4.3}) lead directly to the result.
\ep

\bl\label{lem:4.4}{\rm{(Parameter transformation: cf. \cite{Z}, Lemma 4.)}} Suppose we have
a family of PDFs $f_{\ueta}(\bx)$ parameterized by a $1\times m^\prime$ vector
$\ueta =(\eta_1,\ldots ,\eta_{m^\prime})\in \bbR^{m^\prime}$. Suppose that vector $\ueta$ is a
function of $\utheta\in\bbR^m$. Then
\be \label{eq:4.5}\tJ^\rw_\phi(\BX;\utheta)=\left(\frac{\partial \ueta}{\partial \utheta}\right)^\rT
\tJ^\rw_\phi\Big(\BX;\ueta(\utheta)\Big)
 \left(\frac{\partial \ueta}{\partial \utheta}\right),\ee
with an $\;m^\prime\times m\;$ matrix $\;\diy\frac{\partial \ueta}{\partial \utheta}
=\left(\frac{\partial \eta_i}{\partial  \theta_j}\right)$, $1\leq i\leq m^\prime$, $1\leq j\leq m$.

In the linear case where $\ueta ( \utheta )=\utheta {\tt Q}$ for some $m\times m^\prime$ matrix
$\tt Q$, we obtain:
\be \tJ^\rw_\phi(\BX;\utheta)={\tt Q}\tJ^\rw_\phi\big(\BX;\ueta ( \utheta )\big){\tt Q}^\rT. \ee
\el

\bp Formula (\ref{eq:4.5}) becomes straightforward by substituting the expression
\be \frac{\partial \log f_{\ueta(\utheta)}(x)}{\partial \utheta}=\bigg(\frac{\partial\ueta(\utheta)}{\partial \utheta}\bigg)
\bigg(\frac{\partial \log f_{\ueta}(x)}{\partial \ueta}\bigg)^\rT.\ee
\ep

\def\rp{{\rm p}} \def\rq{{\rm q}}

Concluding this section, we consider a linear model where the parameter is related to an additive shift. Suppose a random vector
$\BX$ in $\bbR^n$ has a joint PDF $f_{\BX}$ and $\bx\in\bbR^n\mapsto\phi (\bx)$ is a given WF. Set:
\be \label{def:WFM}\tL^{\rm w}_\phi (\BX):=\int \frac{\phi(\bx)}{f(\bx)}
\Big(\nabla f(\bx)\Big)^\rT \nabla f(\bx) \rd \bx.\ee
Here and below, we use symbol $\nabla$ for the spatial gradient $1\times n$ vectors as opposite to parameter gradients
$\diy\frac{\partial}{\partial\utheta}$ and $\diy\frac{\partial}{\partial\ueta}$.

Furthermore, set
\be\label{model.linear}\BX=\utheta \tQ+\BY \tP.\ee
Here $\tQ$ and $\tP$ are two matrices, of sizes $m\times n$ and $k\times n$  respectively, with $m\leq k\leq n$.
\def\compl{{\complement}} Next, $\BX\in \bbR^n$ and $\BY\in \bbR^k$.
Let $\bx \in \bbR^n \mapsto \phi(\bx)\geq 0$ be a given WF and set
$$\psi \label{def:psi}(\by )=\psi_\tP(\by)=\int_{\bbR^{n-k}}\phi(\bx ){\mathbf 1}(\bx\tP^\rT =\by) f_{\BX|\BX\tP^\rT}(\bx|\by)\rd\bx^\compl_\tP,\;
\by\in \bbR^n\tP^\rT,$$
where $\bx^\compl_\tP$ stands for the complementary variable in $\bx$, given that $\bx\tP^\rT =\by$.
In Lemma \ref{lem:4.5,6} we present relationships between $\tJ^\rw_\phi(\BX;\utheta)$, $\tJ^{\rm w}_\psi(\BY;\utheta)$,
$\tL^{\rm w}_\phi(\BX)$ and
$\tL^{\rm w}_\psi (\BX\tP^\rT)$ for the above model. (The proofs are straightforward and omitted.)

\bl\label{lem:4.5,6}{\rm{(Cf. \cite{Z}, Lemmas 5 and 6.)}} Assume the model \eqref{model.linear}. Then
\be \tJ^\rw_\phi(\BX;\utheta)=\diy \tQ\tL^{\rm w}_\phi(\BX)\tQ^\rT, \;\;\;\diy\tJ^{\rm w}_\psi (\BY;\utheta)
=\diy \tQ\tP^\rT\tL^{\rm w}_\psi (\BX\tP^\rT) \tP\tQ^\rT, \;\;\;\hbox{and}\;\;\;\tJ^{\rm w}_{\phi}(\BX;\utheta)\geq \diy\tJ^{\rm w}_{\psi}(\BY;\utheta).\ee
\el

\bc\label{cor:4.1}{\rm{(Cf. \cite{Z}, Corollary 1.)}} Let $\tP$ be an $m\times m$ matrix.
Let $\BX$ be a random vector in $\bbR^m$ and WFs $\phi$ and $\psi =\psi_\tP$ be as above. Then
\begin{itemize}
  \item {\rm{(i)}} $\tJ^{\rm w}_{\phi}(\BX;\utheta)\geq \tP^\rT\tJ^{\rm w}_{\psi}(\BX\tP^\rT;\utheta)\tP.$
  \item {\rm{(ii)}} For $\tP$ with orthonormal rows (i.e., with $\tP\tP^\rT$ equal to $I_m$, the unit
$m\times m$ matrix),
\be \label{Cor:(ii)}\tJ^{\rm w}_{\psi}(\BX\tP^\rT; \utheta)\leq \tP^\rT \tJ^{\rm w}_{\phi}(\BX; \utheta)\tP.\ee
  \item {\rm{(iii)}} For $\tP$ with a full row rank $m$, and $\BX\in \bbR^m$ with nonsingular $\tJ^\rw_\phi$,
\be \tJ^{\rm w}_{\psi}(\BX\tP^\rT)\leq \bigg(\tP^\rT {\tJ^{\rm w}_{\psi}(\BX; \utheta)}^{-1} \tP\bigg)^{-1}.\ee
\end{itemize}
\ec

\section{Weighted Cram\'{e}r-Rao and Kullback inequalities}

We start with  multivariate weighted Cram\'{e}r-Rao inequalities (WCRIs). As usually,
consider a family of PDFs $f_{\utheta}(\bx)$,
$\bx\in\bbR^n$, dependent on a parameter $\utheta\in\bbR^m$ and let $\BX=\BX_\utheta$ denote the random vector with PDF $f_{\utheta}$. Let a statistic $\bx\mapsto \BT (\bx)=(T_1(\bx), ..., T_s (\bx))$ and a WF $\bx\mapsto\phi (\bx )\geq 0$ be given. With $\bbE_\utheta$
standing for the expectation relative to $f_{\utheta}$, set:
\be \label{eq:alpha1} \alpha (\utheta )=\bbE_\utheta\,\phi (\BX),\;\;\;\ueta (\utheta )=\bbE_\utheta\,\big[\phi (\BX )\BT (\BX)\big].
\ee
We also suppose that the operations of taking expectation and the gradient are interchangeable:
\be \label{eq:T5.1} \bbE_\utheta \bigg[\phi (\BX)\BS (\BX ,\utheta)\bigg]= \frac{\partial\alpha (\utheta)}{
\partial \utheta},\;\;\;\bbE_\utheta \bigg[\phi (\BX)
\BT(\BX)^\rT\BS (\BX ,\utheta)\bigg]=\frac{\partial\ueta (\utheta)}{\partial\utheta},\ee
assuming C$^1$-dependence in $\utheta\mapsto\alpha (\utheta )$ and $\utheta\mapsto\ueta (\utheta )$ and
absolute convergence of the integrals involved.
Let $\tC^\rw_\phi (\utheta )$ denote the weighted covariance matrix for $\BX$:
\be \label{eq:C}\tC^\rw_\phi (\utheta )=\bbE_\utheta\Big\{\phi (\BX)\Big[\BT (\BX )-\ueta (\BX )\Big]^\rT\Big[\BT (\BX) - \ueta(\BX )\Big] \Big\}
\ee
and $\tJ^\rw_\phi (\BX; \utheta )=\bbE\left[\phi (\BX)\BS (\BX,\utheta )^\rT \BS (\BX,\utheta )\right]$ be
the weighted Fisher information matrix under the WF $\phi$; cf. Eqn \eqref{eq:WtFI}.

\bt \label{thm:5.5}{\rm{(A weighted Cram\'{e}r-Rao inequality, version I; \cite{CT2}, Theorem 11.10.1, \cite{DCT}, Theorem 20.)}} Assuming
$\tJ^\rw_\phi (\BX; \utheta )$ is invertible and under condition \eqref{eq:T5.1},
vectors $\ueta (\utheta )$, $\diy\frac{\partial\alpha (\utheta )}{\partial\utheta}$ and
matrices $\tC^\rw_\phi (\utheta )$, $\tJ^\rw_\phi (\BX; \utheta)$, $\diy\frac{\partial\ueta (\utheta)}{
\partial\utheta}$ obey
\be  \label{WRC}
\tC^\rw_\phi (\utheta )\geq \left[\frac{\partial\ueta (\utheta )}{\partial\utheta}-\big(\ueta (\utheta )\big)^\rT\frac{\partial\alpha
(\utheta )}{\partial\utheta} \right]\,\Big[\tJ^\rw_\phi (\BX; \utheta )\Big]^{-1}\left[
\frac{\partial\ueta (\utheta )}{\partial\utheta}-\big(\ueta (\utheta )\big)^\rT \frac{\partial\alpha(\utheta )}{
\partial\utheta}\right]^\rT.
\ee
\et

\def\BC{\mathbf{C}}
\def\umu{\underline{\mu}}
\def\ulam{\underline{\lambda}}

\bp We start with a simplified version where $s=1$ and $\BT (\BX)=T(\BX )$ and
$\ueta (\utheta )=\eta (\utheta )$
are scalars, keeping general $n,m\geq 1$. By using \eqref{eq:T5.1}, write:
\be \label{eq:T5.11}\begin{array}{l}
\bbE_\utheta\,\Big\{\phi (\BX )\big[T(\BX)-\eta (\utheta )\big]\BS(\BX;\utheta )\Big\}\\
\qquad =
\bbE_\utheta\,\big[\phi (\BX)T(\BX)\BS(\BX;\utheta )\big]-\eta (\utheta )\bbE_\utheta\,\big[\phi (\BX )\BS(\BX;\utheta )\big]
=\diy\frac{\partial\eta (\utheta)}{\partial\utheta}-\eta (\utheta ) \frac{\partial\alpha (\utheta)}{\partial \utheta}.
\end{array}\ee
Then for any $1\times m$ vector $\umu\in\bbR^m$,
\be \label{eq:T5.12}\begin{array}{l}0\leq\bbE_\utheta\,\left\{\phi (\BX)\Big[T(\BX )-\eta (\utheta )-
\BS (\BX,\utheta )\umu^\rT\Big]^2\right\}\\
\qquad\diy =\bbE_\utheta\,\left\{\phi (\BX)\Big[T(\BX )-\eta (\utheta )\Big]^2\right\}+\umu\tJ^\rw_\phi
(\BX;\utheta )\umu^\rT
-2\umu\left(\frac{\partial\eta (\utheta)}{\partial\utheta}-\eta (\utheta ) \frac{\partial\alpha (\utheta)}{
\partial \utheta}\right)^\rT.\end{array}\ee
Taking $\diy\umu =\left(\frac{\partial\eta (\utheta)}{
\partial\utheta}-\eta (\utheta )
\frac{\partial\alpha (\utheta)}{\partial \utheta}\right)\left[\tJ^\rw_\phi (\BX; \utheta )\right]^{-1}$ (which is the minimiser for the RHS in \eqref{eq:T5.12}),
we obtain
\be \label{eq:T5.13}\begin{array}{l}
{\rm{Var}}^\rw_\phi [T(\BX )]:=\bbE_\utheta\,\left\{\phi (\BX)\Big(T(\BX )-\eta (\utheta )\Big)^2\right\}\\
\qquad\diy\geq\left(\frac{\partial\eta (\utheta)}{\partial\utheta}-\eta (\utheta )
\frac{\partial\alpha (\utheta)}{\partial \utheta}\right)\left[\tJ^\rw_\phi (\BX; \utheta )\right]^{-1}\left(\frac{
\partial\eta (\utheta)}{
\partial\utheta}-\eta (\utheta ) \frac{\partial\alpha (\utheta)}{\partial \utheta}\right)^\rT.\end{array}\ee

Turning to the general case $s\geq 1$, set: $T(\BX )=\BT (\BX)\ulam^\rT$ where $1\times s$ vector $\ulam\in\bbR^s$.
Then \eqref{eq:T5.13} yields that for all $\ulam$,
$$\begin{array}{l}\ulam\tC^\rw_\phi (\utheta )\ulam^\rT =
{\rm{Var}}^\rw_\phi [\BT(\BX )\ulam^\rT]:=\bbE_\utheta\,\left\{\phi (\BX)\Big[\BT(\BX )\ulam^\rT-\ueta (\utheta )\ulam^\rT\Big]^2\right\}\\
\qquad\diy\geq\ulam\left(\frac{\partial\ueta (\utheta )}{\partial\utheta}-\big(\ueta (\utheta )\big)^\rT \frac{\partial\alpha(\utheta )}{\partial\utheta} \right)\,\Big[\tJ^\rw_\phi (\BX;\utheta )\Big]^{-1}\left(\frac{\partial\ueta (\utheta )}{\partial\utheta}-\big(\ueta (\utheta )\big)^\rT \frac{\partial\alpha(\utheta )}{\partial\utheta}\right)^\rT
\ulam^\rT,\end{array}$$
implying \eqref{WRC}.
\ep

\bd
The {\bf calibrated} relative WE $K^{\rm w}_\phi (f||g)$ of $f$ and $g$ with WF $\phi$ is defined by
\be  \label{calKullback}
K^{\rm w}_\phi (f||g)=\int \frac{\phi (\bx)f(\bx)}{\alpha(f)}\log\frac{f(\bx)\alpha(g )}{g(\bx)\alpha(f)}\rd \bx
=\frac{D^\rw_\phi (f\|g)}{\alpha (f)} +\log\frac{\alpha (g)}{\alpha (f)} =D (\wtf \|\wtg).\ee
Here $\wtf$ and $\wtg$ are PDFs produced from $\phi f$ and $\phi g$ after normalizing by $\alpha (f)$ and $\alpha (g)$:
\be  \alpha(f)=\int\phi (\bx)f (\bx)\rd \bx,\;\alpha(g)=\int\phi (\bx)g(\bx)\rd \bx,\; \wtf(\bx )=\frac{\phi (\bx )f(\bx )}{\alpha (f)},\;
\wtg(\bx )=\frac{\phi (\bx )g(\bx )}{\alpha (g)},\ee
and $D (\;\cdot\, \|\;\cdot\,)$ is the standard Kullback--Leibler divergence.
\ed

\bt  \label{thm:WKI1}{\rm{(Weighted Kullback inequalities, cf. \cite{FL}.})}\; For given $\phi$ and $f$, $g$
as above, the following bounds hold true.  First, for $1\times n$ vector $\bzeta$,
\be \label{eq:1WKI} K^{\rm w}_{\phi}(f\|g)\geq \sup\,\Big[\frac{\bfe_{\phi}(f)\bzeta^\rT}{\alpha(f)}+\log \alpha(g)-\log M_g(\bzeta ):\;\bzeta\in\bbR^n\Big],\ee
where
\be \label{eq:qua}\bfe_\phi (f)=\int\phi (\bx )f(\bx )\bx\,\rd\bx ,\;\;  M_g(\bzeta)=\diy\int\phi(\bx)g(\bx)
\left[\exp(\bx\bzeta^\rT)\right]\rd\bx .\ee
Second,
\be \label{eq:2WKI} D^{\rm w}_{\phi}(f\|g)\geq
\sup\,\Big[\bfe_{\phi}(f)\bzeta^\rT:\;\bzeta\in\bbM\Big],\ee
where
\be \label{eq:Mmm}\bbM =\left\{\bzeta:\;\int\phi (\bx )\Big(f(\bx )-g(\bx )[\exp(\bx\bzeta^\rT)]\Big)\rd\bx\geq 0\right\}.\ee
\et

\bp First, given $\bzeta\in\bbR^n$, set $\diy\wtG_{\bzeta}(\bx )=\frac{\phi (\bx )g(\bx)\left[\exp(\bx\bzeta^\rT)\right]}{M_{g}(\bzeta)}$.
Following \eqref{eq:qua} and \eqref{calKullback}, obtain:
\be  \diy K^{\rm w}_\phi(f\|g)=\diy D(\wtf\|\wtG_\bzeta)+\diy\int\wtf(\bx)\log \frac{\wtG_\bzeta(\bx)}{\wtg(\bx)} \rd \bx
\geq\diy\int \wtf(\bx)\log \frac{\alpha(g)[\exp(\bx\bzeta^\rT)]}{ M_{g}( \bzeta)}\rd \bx ;\ee
the bound holds as $D(\wtf\|\wtG_\bzeta)\geq 0$ by the Gibbs inequality for the standard Kullback-Leibler divergence. By
taking the supremum, we arrive at \eqref{eq:1WKI}.

Second, write: $G_\bzeta (\bx )=g(\bx)[\exp(\bx\bzeta^\rT)]$ and
\be  \diy D^{\rm w}_\phi(f\|g)=D^{\rm w}_\phi(f\|G_\bzeta )+\bfe_\phi(f)\bzeta^\rT.\ee
For $\bzeta\in\bbM$, the bound $D^{\rm w}_\phi(f\|G_\bzeta )\geq 0$ holds true (the weighted Gibbs inequality (\ref{thm:1.1})). This yields \eqref{eq:2WKI}. 	
\ep

\def\ueps{\underline{\varepsilon}}

An application of the weighted Kullback's inequality is given in the next theorem where we obtain another version of the weighted Cram\'er-Rao inequality.

\bt  {\rm{(A weighted Cram\'{e}r-Rao inequality, version II; \cite{CT2}, Theorem 11.10.1, \cite{DCT}, Theorem 20.)}}\;  Suppose we have a family of
$1\times n$ random vectors $\BX$, with  PDFs $f_\utheta (\bx )$, $\bx\in\bbR^n$,
 indexed by
$\utheta\in \bbR^m$. Suppose that $\diy\frac{f_ \utheta( \bx) \alpha( \utheta +\ueps )
}{f_{ \utheta+\ueps}( \bx)
\alpha( \utheta )}\to 1$ as $\ueps\to 0$ uniformly in $\bx$.
Let $\bx\mapsto\phi (\bx )$ be a given WF. Denoting, as before, the expectation
relative to $f_\utheta$ by $\bbE_\utheta$, set $\alpha (\utheta )=\bbE_\utheta [\phi (\BX )]$,
$\bfe (\utheta )=\bbE_\utheta [\phi (\BX)\BX]$ and
\be  {\wt\tC}^{\rw}_{\phi}(\utheta)=\frac{1}{\alpha(\utheta)}\bbE_ \utheta\big[\phi(\BX)\BX^\rT\BX\big]
-\bfe (\utheta )^\rT\bfe (\utheta ).\ee
Under the assumptions needed to define matrix $\tJ^\rw_{\phi}(\BX; \utheta)$, then
\be \label{eq:wCR21}
\tJ^\rw_{\phi}(\BX; \utheta)\geq\frac{\partial\bfe (\utheta)}{\partial\utheta}^\rT
\Big[{\wt\tC}^{\rw}_{\phi}(\utheta)\Big]^{-1}\frac{\partial\bfe (\utheta)}{\partial\utheta}
+ \alpha( \utheta)^{-1}\frac{\partial \alpha( \utheta)}{\partial \utheta}^\rT\frac{\partial \alpha( \utheta)}{
 \partial \utheta}.\ee
\et

\bp By definition (\ref{calKullback}), for $\ueps\in\bbR^m$,
\be  K^{\rm w}_\phi (f_{ \utheta +\ueps}||f_ \utheta)=-\int\phi (\bx)
\frac{f_{ \utheta +\ueps}(\bx)}{\alpha( \utheta +\ueps)}
\log\frac{f_{ \utheta }(\bx) \alpha( \utheta+\ueps)}{f_{ \utheta+\ueps} (\bx) \alpha( \utheta )}\rd \bx .\ee

Next, set $M (\utheta ,\bzeta )=\bbE_\utheta\left\{\phi (\BX )[\exp(\BX\bzeta^\rT)]\right\}$ and
\be   \Psi (\utheta ,\ueps )=\sup\,\Big[\bfe (\utheta +\ueps )\bzeta^\rT+\log \alpha(\utheta )-\log M (\utheta ,
\bzeta ):\;\bzeta\in\bbR^n\Big].\ee
Then, owing to Theorem \ref{thm:WKI1}, we obtain:
\be \label{eq:1.proofthm1.4}
\diy K^{\rm w}_{\phi} (f_{ \utheta +\ueps}||f_ \utheta)\geq
\Psi( \utheta ,\ueps).\ee
The LHS of \eqref{eq:1.proofthm1.4} is
\be \begin{array}{l} \diy -\diy
\int \phi( \bx) f_{ \utheta+\ueps}( \bx)
\log \frac{f_{ \utheta}( \bx) \alpha( \utheta+\ueps)}{f_{ \utheta+\ueps}( \bx) \alpha( \utheta )}\rd  \bx
=\int \phi( \bx) f_{ \utheta+\ueps}( \bx)\bigg\{
\Big[1-\frac{f_ \utheta( \bx) \alpha( \utheta +\ueps )}{f_{ \utheta+\ueps}( \bx) \alpha( \utheta )}\Big]\\
\\
\qquad\qquad\qquad\diy+\frac{1}{2} \left[1-\frac{f_ \utheta( \bx) \alpha( \utheta +\ueps)}{f_{ \utheta+\ueps}( \bx) \alpha( \utheta )}
\right]^2+O\bigg(\left[1-\frac{f_ \utheta( \bx) \alpha( \utheta +\ueps)}{f_{ \utheta+\ueps}( \bx)
\alpha( \utheta )}\right]^3\bigg)\bigg\}\rd  \bx.\end{array}\ee
Here we have used the Taylor expansion of $\log (1+z)$. The first-order term disappears:
\be
\diy\int \phi( \bx) f_{ \utheta+\ueps}( \bx)\left[1-\frac{f_ \utheta( \bx) \alpha( \utheta +\ueps)}{f_{ \utheta+\ueps}( \bx) \alpha( \utheta )}\right]\rd  \bx=
 \alpha( \utheta +\ueps)- \alpha( \utheta +\ueps )=0.\ee
Next, for small $\ueps$,
\be \begin{array}{l}
\diy\int \phi( \bx) f_{ \utheta+\ueps}( \bx)\left[1-\frac{f_ \utheta( \bx)
 \alpha( \utheta +\ueps)}{f_{ \utheta+\ueps}( \bx) \alpha( \utheta )}\right]^2\rd  \bx\\
 \qquad\qquad\qquad\diy=\ueps\left[
\tJ^\rw_{\phi}(\BX;\utheta)-\frac{1}{\alpha( \utheta)}
\frac{\partial\alpha( \utheta)}{\partial\utheta}\left(\frac{\partial\alpha( \utheta)}{\partial\utheta}\right)^\rT\right]\ueps^\rT
+o(\|\ueps\|^2).\end{array}\ee

Finally, the remainder
\be \diy\lim\limits_{\ueps \rightarrow 0} \frac{1}{\|\ueps\|^2 }\int \phi( \bx) f_{ \utheta+\ueps}( \bx)O\bigg(\left[1-\frac{f_ \utheta( \bx)
 \alpha( \utheta +\ueps )}{f_{ \utheta+\ueps}( \bx) \alpha( \utheta )}\right]^3\bigg)\rd  \bx=o(\|\ueps\|^2 ).\ee   \def\rT{{\rm T}}

\def\btau{{\mbox{\boldmath${\tau}$}}}
\def\mbt{\mathbf{t}}

For the RHS in \eqref{eq:1.proofthm1.4}, we take the point $\btau$ where the gradient has the form \\
$\diy\nabla_\bzeta\Big[
\bfe (\utheta +\ueps )\bzeta^\rT+\log \alpha(\utheta )-\log M (\utheta ,\bzeta )\Big]\Big|_{\bzeta =\btau}=0$, i.e., $$\bfe (\utheta +\ueps )
=\diy\nabla_\bzeta\log M (\utheta ,\bzeta )\Big|_{\bzeta =\btau}=\frac{1}{M(\utheta ,\btau )}\nabla_\bzeta M(\utheta ,\btau )\Big|_{\bzeta =\btau}.$$ Consider the limit
\be \label{eq:3.proofthm1.4}
\diy\lim\limits_{\ueps \rightarrow 0}\frac{1}{\|\ueps\|^2}
\sup\limits_{\mbt\in \mathbb{R}^n}\bigg\{\mbt^\rT \bmu_\phi( \utheta+\ueps)-\overline{\Psi}(\mbt)\bigg\}.\ee
Here $\overline{\Psi}(\mbt)=\log  \alpha( \utheta)+\log \int f_\utheta(\bx)[\exp(\bx\mbt^\rT)]\rd \bx$ denotes the weighted cumulant-generating function
for PDF ${\wt f}_ \utheta$.
The supremum is attained at a value of $\mbt=\btau=\btau (\ueps )$ where the first derivative of the weighted cumulant-generating function equals
$\nabla_\mbt\overline{\Psi}(\mbt=\btau) = \bmu_{\phi}( \utheta+\ueps)$. Here $\bmu_{\phi}( \utheta)=\bbE_ \utheta[\BX \phi(\BX)]/\bbE_ \utheta \phi(\BX)$. We have also $\nabla_\mbt\overline{\Psi}(0)=\bmu_{\phi}( \utheta)$, and therefore the Hessian
\be \label{eq:2.proofthm1.4} \nabla_\mbt\overline{\Psi}(0)=\frac{\partial}{\partial \utheta}\bmu_{\phi}( \utheta) \lim\limits_{\ueps \rightarrow 0}\frac{\partial\ueps}{\partial\btau}.\ee
It also can be seen that
\be  \nabla\nabla\overline{\Psi}(0)=\frac{\bbE_ \utheta\big[\BX^\rT\BX\phi(\BX)\big]}{\bbE_ \utheta [\phi(\BX)]}
-\bmu_{\phi}( \utheta)^\rT\bmu_{\phi}(\utheta):=\overline{\BV}_{\phi}(\BX;\utheta).\ee
In addition, by using the Taylor formula at an intermediate point between $\utheta$ and $\utheta +\ueps$,
\be
\lim\limits_{\ueps \rightarrow 0}\frac{1}{\|\ueps\|^2}\bigg\{\btau^\rT \bmu_{\phi}
( \utheta+\ueps )-\overline{\Psi}(\btau)\bigg\}
=\bigg(\frac{\partial}{\partial \utheta}
\bmu_{\phi}( \utheta)\bigg)\diy\frac{1}{2}[\nabla \nabla \overline{\Psi}(0)]^{-1}\bigg(\frac{\partial}{\partial \utheta}
\bmu_{\phi}( \utheta)\bigg)^\rT.
\ee
Now let us back to the RHS of (\ref{eq:3.proofthm1.4}) which becomes:
\be \label{eq:4.proofthm1.4}
\lim\limits_{\ueps \rightarrow 0}\frac{1}{\|\ueps\|^2}\bigg[
\btau^\rT\bmu_\phi( \utheta+\ueps )-\Psi(\btau)+\log  \alpha( \utheta )\bigg]
= \frac{1}{2}\bigg(\frac{\partial}{\partial \utheta}\bmu_{\phi}( \utheta)\bigg)
[\overline{\BV}_\phi(\BX;\utheta)]^{-1}\bigg(\frac{\partial}{\partial  \utheta}\bmu_{\phi}( \utheta)\bigg)^\rT.
\ee
Now (\ref{eq:4.proofthm1.4}) gives the required result (\ref{eq:wCR21}).
\ep

\br
When $\phi(\bx) \equiv 1$ then $\alpha (\utheta )=1$, $\bfe (\utheta )=\bbE_\utheta\BX$, $\tC^{\rm w}_\phi (\utheta )={\wt\tC}^{\rw}_\phi (\utheta)$, and the two inequalities \eqref{WRC} and \eqref{eq:wCR21} coincide.

In general, these inequalities competing; the question which inequality is stronger is not discussed in this paper. We also note that both inequalities \eqref{WRC} and \eqref{eq:wCR21} lack a covariant property: multiplying WF $\phi$ by a constant has a different impact on the left- and righ-hand sides.
\er



\vskip 10 pt

\noindent
{\bf Acknowledgement}
\vskip 10pt
YS thanks the Office of the Rector, University of Sao Paulo (USP) for the financial support
during the academic year 2013-4. YS thanks Math Department, Penn State University, USA for the hospitality and support during the academic years 2014-6. IS is supported by FAPESP Grant - process number 11/51845-5, and expresses her gratitude to IMS, University of S\~{a}o Paulo, Brazil, and to Math Department, University of Denver, USA for the warm hospitality. SYS thanks the CAPES PNPD-UFSCAR Foundation for the financial support in the year 2014. SYS thanks the Federal University of Sao Carlos, Department of Statistics, for hospitality during the year 2014. MK thanks the Higher School of Economics for the support in the framework of the Global Competitiveness Program.

\noindent Yuri Suhov: DPMMS, University of Cambridge, UK; Math Dept, Penn State University, PA, USA; IPIT RAS, Moscow, RF
\vskip 10pt

\noindent Izabella Stuhl: IMS, University of S\~{a}o Paulo, Brazil; Math Dept, University of Denver, CO, USA; University of Debrecen, Hungary
\vskip 10pt

\noindent Salimeh Yasaei Sekeh: Stat Dept, Federal University of S$\tilde{\rm a}$o Carlos, SP, Brazil
\vskip 10pt

\noindent Mark Kelbert: Math Dept, University of Swansea, UK; Moscow Higher School of Economics, RF

\end{document}